\documentclass[sigconf, screen=true, printacmref=false, printccs=false, printfolios=false]{acmart}
\setcopyright{none}

\usepackage{xcolor}
\definecolor{thedarkblue}{RGB}{0,0,120} 
\definecolor{mydarkblue}{rgb}{0,0.08,0.45} 
\definecolor{darkblue}{rgb}{0,0.08,180}
\colorlet{TufteRed}{red!80!black}
\definecolor{theblue}{RGB}{0,0,180}
\colorlet{thered}{TufteRed}
      
\usepackage{microtype}
\usepackage{balance}

\usepackage{booktabs}
\usepackage{tabularx}

\usepackage{amsmath,amsthm}

\newcommand{\eat}[1]{\ignorespaces}
\usepackage{comment}

\newcommand{\journal}[1]{} 

\usepackage{tikz}
\usepackage{verbatim}
\usetikzlibrary{arrows}
\usetikzlibrary{shapes,snakes}
\usetikzlibrary{decorations.pathmorphing} 
\usetikzlibrary{fit}					
\usetikzlibrary{backgrounds}	

\usepackage{ragged2e}
\usepackage{multirow}
\usepackage{microtype}
\usepackage{balance}
\usepackage{setspace}

\graphicspath{{./}{./graphics/}}
\newcolumntype{H}{>{\setbox0=\hbox\bgroup}c<{\egroup}@{}}

\newcolumntype{R}[1]{>{\RaggedLeft\arraybackslash}} 
\newcolumntype{L}[1]{>{\RaggedRight\arraybackslash}} 

\newcommand{\eg}{\emph{e.g.}}

\newtheorem{Definition}{\hspace{-1em}\bfseries{Definition}}

\AtBeginEnvironment{pmatrix}{\setlength{\arraycolsep}{2pt}}

\providecommand{\mat}[1]{\boldsymbol{\mathrm{#1}}}%
\renewcommand{\vec}[1]{\boldsymbol{\mathrm{#1}}}

\DeclareMathOperator{\hugeE}{\mbox{\huge\raise-0.3ex\hbox{E}}}
\DeclareMathOperator{\p}{\mathbb{P}}
\DeclareMathOperator{\hugep}{\mbox{\huge\raise-0.3ex\hbox{$\p$}}}

\newcommand{\RR}{\mathbb{R}}

\providecommand{\mH}{\ensuremath{\mat{H}}}

\providecommand{\mW}{\ensuremath{\mat{W}}}

\providecommand{\vc}{\ensuremath{\vec{c}}}

\providecommand{\vh}{\ensuremath{\vec{h}}}

\providecommand{\vn}{\ensuremath{\vec{n}}}

\providecommand{\vp}{\ensuremath{\vec{p}}}

\providecommand{\vs}{\ensuremath{\vec{s}}}

\providecommand{\vy}{\ensuremath{\vec{y}}}
\providecommand{\vz}{\ensuremath{\vec{z}}}

\usepackage{algorithm}
\usepackage{algorithmicx}
\usepackage{algpseudocode}
\algblockdefx[parallel]{ParFor}{EndPar}[1][]{$\textbf{parallel for}$ #1 $\textbf{do}$}{$\textbf{end parallel}$}
\algrenewcommand{\alglinenumber}[1]{\fontsize{6.5}{7}\selectfont#1}
\algtext*{EndPar}

\algblockdefx[parallel]{parfor}{endpar}[1][]{$\textbf{parallel for}$ #1 $\textbf{do}$}{$\textbf{end parallel}$}
\algrenewcommand{\alglinenumber}[1]{\scriptsize#1:}

\algblockdefx[parallel]{parallelfor}{parallelend}
[1][]{\textbf{parallel for} #1}
{\textbf{end parallel}}

\usepackage{nicefrac}

\algnotext{EndIf}
\algnotext{EndProcedure}

\usepackage{subfigure}
\usepackage{graphbox}
\usepackage{svg}
\usepackage{nicefrac}

\usepackage[toc,page]{appendix}

\DeclareMathAlphabet{\mathbcal}{OMS}{cmsy}{b}{n}
\usepackage{mathrsfs}
\usepackage{comment}

\usepackage{bm}
\usepackage{bbm}
\usepackage{enumitem}
\AtBeginDocument{
  \providecommand\BibTeX{{
    \normalfont B\kern-0.5em{\scshape i\kern-0.25em b}\kern-0.8em\TeX}}}

\usepackage{xargs}      
\usepackage{soul}       
\usepackage{color}      

\definecolor{lightpink}{RGB}{237,157,202}
\definecolor{lightred}{RGB}{210,121,121}
\definecolor{lightorange}{RGB}{230,170,50}
\definecolor{lightgold}{RGB}{210,194,121}
\definecolor{lightgreen}{RGB}{121,210,121}
\definecolor{lightaqua}{RGB}{121,206,210}
\definecolor{lightblue}{RGB}{121,124,210}
\definecolor{lightpurple}{RGB}{153,102,255}
\definecolor{figureRed}{RGB}{214,39,40}
\definecolor{figureBlue}{RGB}{31,119,180}

\begin{document}

\title{Graph Learning with Localized Neighborhood Fairness}

\settopmatter{authorsperrow=5}

\author{April Chen}
\affiliation{%
  \institution{Harvard University}
}

\author{Ryan A. Rossi}
\affiliation{
  \institution{Adobe Research}
}

\author{Jane Hoffswell}
\affiliation{
  \institution{Adobe Research}
}

\author{Shunan Guo}
\affiliation{
  \institution{Adobe Research}
}

\author{Sungchul Kim}
\affiliation{
  \institution{Adobe Research}
}

\author{Nedim Lipka}
\affiliation{
  \institution{Adobe Research}
}

\author{Gromit Chan}
\affiliation{
  \institution{Adobe Research}
}

\author{Eunyee Koh}
\affiliation{%
  \institution{Adobe Research}
}

\author{\text{Nesreen K. Ahmed}}
\affiliation{%
  \institution{Intel Labs}
}
\email{}

\renewcommand{\shortauthors}{A. Chen et al.}

\begin{abstract}
Learning fair graph representations for downstream applications is becoming increasingly important,
but existing work has mostly focused on improving fairness at the global level by either modifying the graph structure or objective function without taking into account the local neighborhood of a node.
In this work, we formally introduce the notion of neighborhood fairness and develop a computational framework for learning such locally fair embeddings.
We argue that the notion of neighborhood fairness is more appropriate since GNN-based models operate at the local neighborhood level of a node.
Our neighborhood fairness framework has two main components that are flexible for learning fair graph representations from arbitrary data:
the first aims to construct fair neighborhoods for any arbitrary node in a graph and
the second enables adaption of these fair neighborhoods to better capture certain application or data-dependent constraints, such as allowing neighborhoods to be more biased towards certain attributes or neighbors in the graph.
Furthermore, while link prediction has been extensively studied, we are the first to investigate the graph representation learning task of fair link classification.
We demonstrate the effectiveness of the proposed neighborhood fairness framework for a variety of graph machine learning tasks including fair link prediction, link classification, and learning fair graph embeddings.
Notably, our approach achieves not only better fairness but also increases the accuracy in the majority of cases across a wide variety of graphs, problem settings, and metrics.

\end{abstract}



\maketitle

\section{Introduction}

\begin{figure}[h!]
\centering
\vspace{-2mm}
\includegraphics[width=\linewidth]{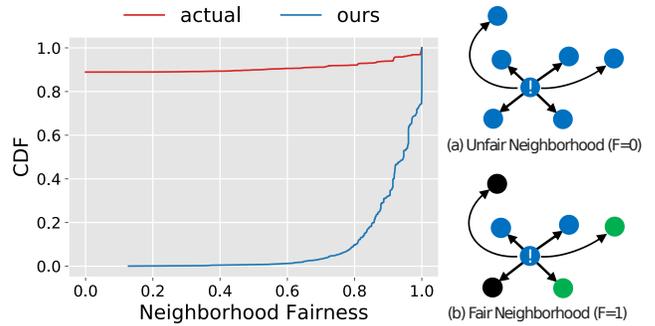}
\vspace{-6mm}
\caption{%
Most neighborhoods in real-world (\textcolor{figureRed}{actual}) graphs are \emph{unfair}.
However, our approach (\textcolor{figureBlue}{ours}) derives fair neighborhoods and can be leveraged for learning fair node embeddings in downstream tasks.
For instance, 90\% of neighborhoods in Cora are maximally unfair~(a) with a neighborhood fairness score of 0, whereas only a handful achieve the max fairness score~(b).
Conversely, our approach derives fair neighborhoods with none of them reaching the max unfairness score; in fact, only about 10\% of the neighborhoods derived using our approach have a fairness score $<$ 0.8. 
}
\label{fig:neigh-fairness-teaser}
\vspace{-2mm}
\end{figure}

Neighborhoods are at the heart of all graph neural network (GNN) methods~\cite{scarselli2008graph,kipf2016semi,li2018adaptive,wu2019simplifying,velivckovic2017graph,rossi2018deep,zhang2019heterogeneous} and therefore fundamentally important for ensuring fairness in GNNs. 
Notably, since all GNNs rely strictly on the neighborhoods of nodes in the graph, the fairness is fundamentally tied to the neighborhoods that are used to train them.
Hence, if a neighborhood is unfair, the resulting embeddings will also be unfair.
Despite this fact, existing work mainly ignores the fairness of individual neighborhoods~\cite{FairDrop,pmlr-v119-buyl20a,current2022fairmod,loveland2022fairedit,FairAdj,DBLP:journals/corr/abs-2104-14210}.

To address this, we first introduce the notion of \emph{neighborhood fairness}.
Intuitively, an unfair neighborhood of a node is one where all neighbors have the same sensitive attribute (Figure~\ref{fig:neigh-fairness-teaser}a), and conversely, the neighborhood that is most fair is one where the sensitive attribute of the neighbors are uniformly distributed among them (Figure~\ref{fig:neigh-fairness-teaser}b). 
Strikingly, neighborhoods in real-world graphs often exhibit extreme levels of unfairness, where all neighbors share the same sensitive attribute.
For example, in Figure~\ref{fig:neigh-fairness-teaser},
roughly 90\% of neighborhoods in Cora have the maximum unfairness $\mathbb{F}(\vp_i)=0$; other graphs demonstrate similarly dramatic levels of unfairness. See Appendix~\ref{sec:neigh-fairness-metric} for further details.
This fact motivates the importance of the proposed notion of neighborhood fairness and its impact on the unfairness of all GNN-based methods due to training via neighborhood aggregation.

\begin{figure}[t!]
\vspace{-6mm}
\centering

\hspace{-4mm}\includegraphics[width=1.05\linewidth]{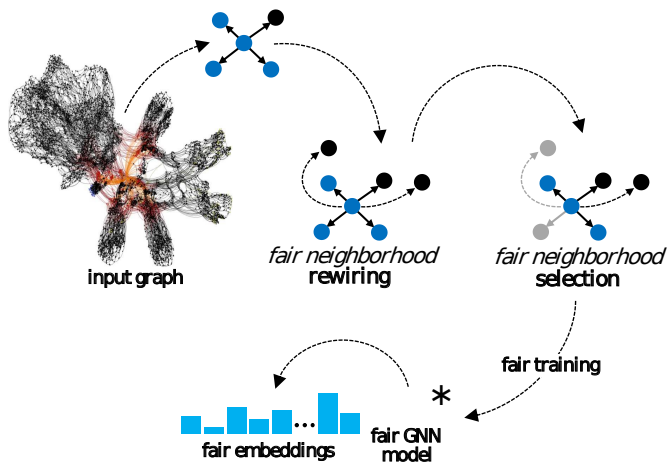}

\vspace{-3mm}
\caption{
Overview of the neighborhood fairness graph representation learning framework.
For simplicity, only a single neighborhood is shown, however, this process is applied to all neighborhoods, and can even be repeated for every epoch during training.
 }
 \vspace{-4mm}
\label{fig:overview}
\end{figure}

Based on these principled notions of neighborhood fairness, we develop a framework for incorporating neighborhood fairness into any existing GNN-based model. 
By ensuring the fairness of neighborhoods (and consequently the fairness of the neighborhood-level aggregation for each node) on which the GNNs are trained,
their results correspondingly become more fair. 
The framework gives rise to fair GNN models that can then be leveraged for a variety of important applications including link prediction, recommendation, link classification, or simply to learn embeddings of the nodes, which can then be used for a variety of real-world tasks.
Our neighborhood fairness framework, \textsc{FairNeigh}, has two main components (Figure~\ref{fig:overview}): 
a fair neighborhood \emph{rewiring} component and a fair neighborhood \emph{selection} component.
In the first component, we rewire or edit the neighborhood of a node by adding or removing edges, to ensure it is fair with respect to the proposed notions of neighborhood fairness. Depending on the task for which the embeddings are used, one can flexibly employ different methods to choose these added and removed edges, such that both accuracy and fairness can be improved. For instance, for embeddings used in link prediction, one can choose edges to fulfill fairness requirements such that the edges are also similar to existing edges, thus having a high likelihood of being in the test set and improving accuracy.
In the second component, we use randomized selection to control 
(i) the homophily and heterophily of constructed edges, 
(ii) the ratio of original to constructed edges in the final neighborhoods, and
(iii) the overall heterophily.
This framework provides several levers for the user to control properties of the fair neighborhoods in a flexible and task-dependent manner. 
After the randomized selection step, the fair neighborhoods are used to learn fair and accurate embeddings. 
Our framework is highly flexible and generalizable, allowing the user to not only improve fairness, but also accuracy.

Through extensive experiments, we demonstrate the effectiveness of the notion of neighborhood fairness and the proposed computational framework for learning GNN-based models that preserve neighborhood fairness.
These experiments demonstrate the effectiveness in terms of both accuracy and fairness on a wide variety of graphs 
and for several important tasks including link prediction, link classification, and fair graph embeddings.

\medskip\noindent\textbf{Summary of Main Contributions.} 
The key contributions of this work are as follows:
\begin{itemize}[leftmargin=*]
    \item \textbf{Neighborhood Fairness: } 
    Since the fairness of all GNN-based methods depend on the underlying neighborhoods used during training, we formally introduce the notion of neighborhood fairness and a relaxation of this stricter notion.
    Most importantly, 
    neighborhood fairness 
    enables one to develop techniques that capture fairness at the localized neighborhood level of the nodes in the graph, which 
    is of fundamental importance since all GNN methods and the resulting fairness of these models depend on the neighborhoods used to train them.

    \item \textbf{Computational Framework: } We contribute an algorithmic framework, \textsc{FairNeigh}, for training GNN models and learning embeddings that are inherently fair with respect to the notions of neighborhood fairness. 
    The neighborhood fairness framework is highly flexible with many interchangeable components 
    that can be fine-tuned for specific downstream applications.
   
    \item \textbf{Fairness Analysis: } We perform an analysis of neighborhoods and demonstrate the extent of unfairness in local neighborhoods from real-world graphs.
    We also introduce a corresponding metric for quantifying neighborhood fairness.

    \item \textbf{Fair Link Classification: } 
    To the best of our knowledge, this work is the first to study fair link classification.

    \item \textbf{Effectiveness: } Through extensive experiments, we demonstrate the effectiveness of neighborhood fairness across a wide variety of graphs and learning tasks including fair link prediction, link classification, and learning fair graph embeddings.
  
\end{itemize}

\vspace{-3mm}

\section{Related Work} \label{sec:related-work}
Graph embeddings are used in a variety of machine learning tasks, including relation prediction, clustering, and node classification~\cite{https://doi.org/10.48550/arxiv.1709.05584}.
Some work has begun to explore fairness and bias in graph embeddings. 
Bose~et~al.~\cite{bose2019compositional} incorporated compositional fairness constraints into graph embeddings, building on an approach by Edwards and Storkey~\cite{https://doi.org/10.48550/arxiv.1511.05897}. Buyl~et~al.~\cite{pmlr-v119-buyl20a} propose a method to de-bias embeddings from a biased prior. Spinelli~et~al.~\cite{DBLP:journals/corr/abs-2104-14210} propose an edge-dropout algorithm to improve fairness and reduce homophily in graph representation learning. Prior work has also explored debiasing knowledge graph embeddings~\cite{fisher2020debiasing, arduini2020adversarial}. Many papers have tried to create fair embeddings for GNNs~\cite{palowitch2019monet, creager2019flexibly, wang2022unbiased, ma2022learning, dong2022edits}, while
others work directly on the model~\cite{agarwal2021towards, dai2021say, rahman2019fairwalk, buyl2021kl}.
Recent work called FairAdj~\cite{FairAdj} aims to learn a fair adjacency matrix for a downstream link prediction task.
More recently, optimal transport techniques were used to obtain dyadic fairness~\cite{yang2022obtaining} and to create an embedding-agnostic algorithm to repair adjacency matrices for fair link prediction~\cite{laclau2021all}.

A variety of other work has also explored the topic of link prediction. Tsioutsiouliklis~et~al.~\cite{tsioutsiouliklis2021fairness} find algorithms for fairness in the PageRank algorithm. Kansal~et~al.~\cite{kansal2022flib} propose Fair Link Prediction in Bipartite Networks.
Masrour~et~al.~\cite{masrour2020bursting} address the problem of link prediction homophily with postprocessing. Recent work has also exploited communities to obtain fair link predictions in complex networks~\cite{saxena2021hm}.
More recently, Sean~et~al.~\cite{current2022fairmod} studied graph modification strategies that perform edits to the input graph during GNN training for downstream link prediction. Some of these link prediction problems focus on bias in recommender systems~\cite{wu2021learning, li2021towards, li2022fairsr}.
Prior work has also explored debiasing clustering models~\mbox{\cite{kleindessner2019guarantees, gupta2021protecting}.} Other related work has identified node degree as a notable source of bias~\cite{tang2020investigating, kang2022rawlsgcn, jiang2022fmp}.
Furthermore, there have been other works on specific applications where fairness is particularly important to guarantee. Some recent work has investigated approaches to fair influence maximization in communities, modeled by graphs ~\cite{rahmattalabi2020fair, farnad2020unifying, khajehnejad2020adversarial}. In earlier work, Rahmattalabi et al.~\cite{rahmattalabi2019exploring} propose a formulation of the robust graph covering problem with group fairness constraints, and analyze the Price of Fairness (PoF). There have also been studies of ranking-based graph fairness ~\cite{dong2021individual, krasanakis2020applying}. Kang et al.~\cite{inform2020} propose a definition and mitigation methods for individual fairness in graph mining. 
Recent work has also focused on adaptive data augmentations to learn fair node representations~\cite{kose2022fair}.
Subgroup fairness in graphs have also been studied for improving spam detection~\cite{liu2022subgroup}. 
FairEdit~\cite{loveland2022fairedit} leverages both greedy edge additions and deletions to improve fairness in GNNs.
Khajehnejad et al.~\cite{khajehnejad2021crosswalk} propose CrossWalk, a method that enhances fairness in graph algorithms by biasing random walks to cross group boundaries.

While all GNN-based methods are trained based on the node neighborhoods, existing work on fairness in graph representation learning does not focus on the notion of neighborhood fairness.
In this work, we introduce the notion of neighborhood fairness and develop a neighborhood fairness framework that can be leveraged by any GNN-based model or graph representation learning approach.
In addition, we formulate a new fairness metric for graphs based on the notion of neighborhood fairness.
To the best of our knowledge, this work is also the first to study the fair link classification task.

\section{Neighborhood fairness}
\label{sec:neigh-fairness-notions}
Prior work in graph embedding fairness have worked with global graph fairness, which overlooks fairness with respect to each node's local neighborhood. 
However, graph updates are passed through the neighbors; thus, if a node's neighborhood is unfair, it will receive unfair information, and the resulting embedding will be unfair. 
We therefore formally introduce the notion of neighborhood fairness for GNN models.
Let $S$ denote the set of sensitive attribute values and let $s_i$ denote the sensitive attribute value of node $i$. 
We begin by introducing the notion of an attributed neighborhood, which is used as a basis for deriving fair neighborhoods.

\begin{Definition}[Attributed Neighborhood]\label{def:attributed-neighborhood}
Given an arbitrary node $i$ in $G$, the \emph{attributed neighborhood} $N_{i}^{s}$ of node $i$ is the set of nodes with attribute value $s$ that are reachable by following edges originating from $i$ within $1$-hop distance.
More formally, 
\begin{equation}\label{eq:attributed-neighborhood}
N_{i}^{s} = \{ j \in V \, | \, (i,j) \in E \wedge s_i=s \}
\end{equation}\noindent
Intuitively, a node $j \in N_{i}^{s}$ iff there exists an edge $(i,j) \in E$ between $i$ and $j$ and the sensitive attribute value of node $j$ denoted as $s_j$ is $s$.
\end{Definition}\noindent

\noindent Note this notion naturally generalizes to multiple (sensitive) attributes. Further, let $N_i^{s_i}$ be the set of nodes in the neighborhood of $i$ with the same attribute as node $i$, and $N_i^{\overline{s_i}} = N_i \setminus N_i^{s_i}$. We also denote $\Gamma_i$ as the set of distance-two neighbors of $i$.

\noindent With these definitions in mind, we define local neighborhood fairness.
The strictest notion of local neighborhood fairness is when the neighborhood is completely agnostic to the sensitive attribute:

\begin{Definition}[Exact Local Neighborhood Fairness] \label{def:exact-local-fairness}
\hspace{10px}
A node $i$'s neighborhood $N_i$ is \textbf{exactly locally fair} if each sensitive attribute value in the set $S$ is equally represented in the neighborhood.
$$\forall s_j, s_k \in S, |N_i^{s_j}|=|N_i^{s_k}|$$
\end{Definition}
\noindent 
Exact neighborhood fairness implies that node $i$ should receive no information about its own sensitive attribute value from its neighbors.
However, this notion of fairness can be difficult to achieve, especially in real graphs where the distribution of sensitive attributes is very non-uniform.
We thus define a relaxation. 
Intuitively, we achieve fairness when it is ``hard'' to identify node $i$'s sensitive attribute value based on its neighborhood, and when node $i$'s updates are not unfairly skewed towards any particular attribute value:
\begin{Definition}[Relaxed Local Neighborhood Fairness] \label{def:relaxed-local-fairness}
A node $i$'s neighborhood $N_i$ is \textbf{counterfactually locally fair} if node $i$'s sensitive attribute value is equally represented to the most common other attribute value.
$$|N^{s_i}_i| = \text{max}_{s \in S, s\neq s_i}|N^{s}_i|$$
\end{Definition}
\noindent 
``Relaxed local fairness'' and ``counterfactual local fairness'' are used interchangeably.
While this notion is weaker than exact neighborhood fairness, it still intuitively preserves some fairness. In many graphs, most node's neighborhoods are highly skewed towards their own sensitive attribute values; that is, one could achieve high accuracy in predicting node $i$'s sensitive attribute value by just observing its neighborhood. 
If relaxed neighborhood fairness is satisfied, then predicting the most common attribute value in the neighborhood will only succeed at most half of the time.

\newcommand{\setAlgFontSize}{\fontsize{8pt}{9pt}\selectfont} 
\newcommand{\multiline}[1]{\State \parbox[t]{\dimexpr\linewidth-\algorithmicindent}{#1\strut}}
\newcommand{\multilinenospace}[1]{\State \parbox[t]{\dimexpr0.98\linewidth-\algorithmicindent}{\begin{spacing}{1.0}\setAlgFontSize#1\strut\end{spacing}}}
\newcommand{\multilinenospaceD}[1]{\State \parbox[t]{\dimexpr0.96 \linewidth-\algorithmicindent}{\begin{spacing}{1.1}\setAlgFontSize#1\strut \end{spacing}}}

\begin{figure}[t!]
\vspace{-4.5mm}
\algblockdefx[parallel]{parfor}{endpar}[1][]{$\textbf{parallel for}$ #1 $\textbf{do}$}{$\textbf{end parallel}$}
\algrenewcommand{\alglinenumber}[1]{\fontsize{7.5}{8}\selectfont\bf #1:\,\,}
\begin{center}
\begin{algorithm}[H]
\caption{\,
Neighborhood Fairness Framework 
}
\label{alg:framework}
{%
\begin{spacing}{1.2}
\begin{algorithmic}[1]
\Require a graph $G$
\vspace{-0.5mm}
\Ensure fair embeddings $\mH \in \RR^{n \times d}$

\For{$k=1,\ldots,K$}
    \For{$i \in V$}
        \hspace{0.5mm}
        \State $F_i = \texttt{FairNeigh}(N_i, G)$ \Comment{Section~\ref{sec:rewiring}}
        \hspace{8.5mm}
        \State Apply selection functions $F_i^{\prime} = \Pi(F_i)$ \Comment{Section~\ref{sec:neigh-selection-randomization}}

        \State $\vh_{F_i^{\prime}}^{k} = \textsc{Aggr}_k\Big(\big\{\vh_j^{k-1},\;\, \forall j \in F_i^{\prime}\big\}\Big)$ 
        \Comment{Section ~\ref{sec:training}}
        
        \State $\vh_i^{k} \leftarrow \sigma\Big(\mW^k \cdot \Psi\Big(\vh_{F_i^{\prime}}^{k}, \vh_i^{k-1}\Big)\Big)$ 
        
    \EndFor
\EndFor

\State {\bf return} fair learned embeddings $\mH \in \RR^{n \times d}$ \label{algline:return-embedding}
\end{algorithmic}
\end{spacing}
}
\end{algorithm}
\end{center}
\vspace{-4.5mm}
\end{figure}

\section{Framework} \label{sec:framework}

An overview of the neighborhood fairness framework is provided in Figure~\ref{fig:overview}.
Given a graph $G$, we first perform \emph{fair neighborhood rewiring} (Section \ref{sec:rewiring}) to derive fair neighborhoods, then \emph{fair neighborhood selection} (Section~\ref{sec:neigh-selection-randomization}) to control for factors 
such as the proportion of homogeneous vs. heterogeneous edges or original vs. constructed edges. 
Finally, Section~\ref{sec:training} describes training using these neighborhoods.
See Alg.~\ref{alg:framework} for a more technical overview.

\subsection{Fair Neighborhood Rewiring} \label{sec:rewiring}
Now we describe the fair neighborhood rewiring component of our framework and a few techniques to ensure neighborhoods are fair via rewiring (edits) such as adding, deleting or changing the neighbors of a node to preserve some notion of neighborhood fairness. We introduce a general framework in Alg.~\ref{alg:neigh-fairness-rewiring-general} that gives rise to many different fair neighborhood rewiring techniques.
In particular, one can carefully select the various functions given as input to Alg.~\ref{alg:neigh-fairness-rewiring-general} to obtain a new neighborhood rewiring procedure.
Overall, the fair neighborhood rewiring framework has the following key steps:

\begin{enumerate}
    \item \textbf{Fairness function: } We first use some node fairness function $f$, which outputs vector $\vn_i$ which describes how many nodes of each sensitive attribute value should be in the neighborhood for it to be fair (Line 1).
    \item \textbf{Remove nodes: } If there are too many nodes of some sensitive attribute, then function $\mathcal{R}$ reduces the nodes by sampling from the neighborhood, \eg if $N_i$ is primarily blue nodes $\mathcal{R}(N_i, \vn_i)$ might sample some blue nodes to remove, and return the rest of the neighborhood (Line 2).
    \item \textbf{Add nodes: } To supplement the existing neighborhood, we use node expansion function $\xi$, which outputs a set of nodes $W$ from which to sample (e.g. the two-hop neighborhood) (Line 4). Then, for each sensitive attribute value $s$, we filter the set $W$ to $W^s \subset W$, the subset with attribute value $s$ (Line 6). We then sample nodes from $W^s$ based on a distribution $\mathcal{D}$, which can be selected based on the task of interest, and add them to the neighborhood (e.g. selecting nodes based on Jaccard similarity for link prediction) (Line 7-8).
\end{enumerate}

\noindent Importantly, each part of this framework can be selected flexibly depending on the fairness definition and task. For instance, the node fairness function $f$ can output a vector of node numbers that would achieve exact fairness, relaxed fairness, or some other neighborhood fairness definition.
Depending on the task, the node reduction function $\mathcal{R}$ could keep all nodes, or keep only nodes with similar features to node $i$ for a link prediction task. The node expansion function $\xi$ 
could naively output the set of all nodes outside the neighborhood $F_i$, or in the case of link prediction, could output the two-hop neighborhood $\Gamma_i$, which contains nodes that are more likely to be similar to $i$. Similarly, the distribution $\mathcal{D}$ could just be a naive uniform distribution, or could be a weighted distribution where nodes are weighted by a similarity function.

To demonstrate this, we now formally present two specific techniques that use this framework. First, in Section~\ref{sec:rewiring-task-indep-neigh-fairness-exact} we describe a simple task-independent neighborhood fairness rewiring approach that satisfies the exact neighborhood fairness notion proposed in Definition~\ref{def:exact-local-fairness}. 
Then in Section~\ref{sec:rewiring-task-link-neigh-fairness} we describe a link prediction oriented rewiring technique, that  corresponds to the relaxed neighborhood fairness notion introduced  Definition~\ref{def:relaxed-local-fairness}.

\begin{figure}[t!]
\vspace{-2mm}
\algrenewcommand{\alglinenumber}[1]{\fontsize{7.5}{8}\selectfont\bf #1:\,\,}
\begin{center}
\begin{algorithm}[H]
\caption{\,
Neighborhood Fairness Rewiring Framework
}
\label{alg:neigh-fairness-rewiring-general}
{%
\begin{spacing}{1.1}
\begin{algorithmic}[1]
\vspace{1mm}
\Require Neighborhood $N_i$, node $i$, sensitive attribute $\vs$, 
node fairness function $f: \text{node} \to \mathbb{N}^{|S|}$,
reduction function $\mathcal{R}$,
expansion function $\xi$, 
task-dependent (weighted) distribution $\mathcal{D}$ 

\Ensure Fair neighborhood $F_i$ for node $i$
\vspace{1.5mm}

\State Set $\vn_i = f(i)$
\Comment{$n_{is}$ is the number of nodes of label $s$ that should be in the neighborhood of $i$ to ensure fairness} 

\State $F_i \leftarrow \mathcal{R}(N_i, \vn_i)$ \Comment This function removes nodes if based on $\vn_i$ there are too many of attribute $s$
\For{$s=s_1, \ldots, s_{|S|}$}
        \State Set $W = \xi(F_i)$  
    \For{$k=1,\ldots,n_{is}$}
        \State Let $W_s = \{j \in W | s_j = s\}$
        \State Select $j \sim \mathcal{D}(W^s)$ \Comment{Sample nodes via weighted dist.}

        \label{alg:rewiring-dist}
        \State $F_i = F_i \cup \{j\}$
    \EndFor
\EndFor

\vspace{0.5mm}
\end{algorithmic}
\end{spacing}
}
\end{algorithm}
\end{center}
\vspace{-4mm}
\end{figure}

\subsubsection{Exact Task-Independent Neighborhood Fairness Rewiring}\label{sec:rewiring-task-indep-neigh-fairness-exact}

Using this framework, we describe an ``exact'' variant of the task-independent fair neighborhood rewiring that essentially ensures that a neighborhood of a node contains the same number of nodes for every different sensitive attribute value.

In this approach, let $f(i) = \vn_i$ such that for each $i$ and $s$,  $$n_{is} = |N_i^{s_{\max}}| - |N_i^s|$$  where $s_{\max}$ is the most common sensitive attribute value. In other words, for all other sensitive attribute values $s$, we must add nodes of that attribute until $|N_i^s| = |N_i^{s_{\max}}|$. The inclusion of these nodes thus guarantees the notion of exact neighborhood fairness (Definition \ref{def:exact-local-fairness}). We need not remove any nodes, so the reduction function $\mathcal{R}$ is the identity.
We then find nodes elsewhere in the graph to fulfill the nodes needed for each $s$.
Given a sensitive attribute value $s$, 
we define $V_s = \{\,i \in V \;|\; s_i = s\,\}$ and 
choose node expansion function $\xi$ such that $\xi(F_i) =  V_s \setminus F_i$, and we let the distribution function be an arbitrary distribution $\mathcal{D}$; this could be Uniform, WeightedUniform, or follow some other distribution based on the user's requirements.

\subsubsection{Link Prediction Fair Neighborhood Rewiring}\label{sec:rewiring-task-link-neigh-fairness}

This framework can also be used for different definitions of fairness and downstream tasks. For a link prediction task, one can achieve higher accuracy through careful choice of the nodes added to the neighborhood. We now demonstrate a rewiring method that achieves relaxed (counterfactual) fairness and carefully selects nodes that will lead to better training. In particular, we aim to add node $j$ to the neighborhood of $i$ if $i,j$ is likely to be an unobserved edge. These could be nodes that are closely located, or similar based on node features. 

For this fairness function $f$, let it set fair sizes $\vn_i$ for each sensitive attribute value using the notion of counterfactual local neighborhood fairness. In this example, we aim to balance two most common sensitive attribute values in the neighborhood. Let $s_2 = \text{argmax}_s|N_i^{S\setminus{s_{\max}}}|$, then set \begin{align}
    n_{is_2} = |N_i^{s_{\max}}| - |N_i^{{s_2}}|
\end{align}
\noindent and for all other sensitive attribute values $s'$, set $n_{is'} = 0$.
Since the end task is link prediction, we want additional nodes to be ``close" in the graph to node $i$. Thus, we choose $\xi$ so that $W = \xi(F_i) = \Gamma_i \setminus F_i$, or the two-hop neighborhood of $i$, not including one-hop neighbors.For our distribution $D$, we would like to be more likely to select nodes that are more ``similar" to node $i$; thus, we sample from $W^{s_2}$ via a WeightedUniform distribution, where nodes are weighted by some similarity function $\phi_i$, where $\phi_i(j)$ outputs the similarity between $i$ and $j$. 
These similarity functions should weight nodes $j$ highly if they will aid link prediction training.
To this end, many similarity functions are possible and one such approach is:
\begin{align}\phi_i(j) = |N_i \cap N_j|\end{align}
where $\phi_i(j)$ is the number of neighbors $i$ and $j$ have in common. 
This weighting function is simple and intuitive in that it depends only on the structure of the graph. 
However, this weighting function is completely interchangeable; other possibilities include Jaccard similarity or leveraging the distance between embedding vectors of $i$ and $j$ that are currently being learned.
By rewiring the neighborhood of node $i$, we now have a \textit{counterfactually fair neighborhood} as defined in Definition \ref{def:relaxed-local-fairness}, since $|F_i^{s_{\max}}| = |F_i^{s_i}|$.

\subsection{Neighborhood Selection Randomization}\label{sec:neigh-selection-randomization}

Given the fair neighborhoods $F_i$ constructed as described in the previous section, we now introduce levers to control certain properties of the neighborhoods during the learning of node embeddings. 
However, it is useful to be able to control how many edges constructed in Section \ref{sec:rewiring} are used, as well as the relative homophily of these edges and the overall homophily.
The second part of our framework consists of three randomization layers: 
selecting what proportion of constructed edges are homophilous (Section \ref{sec:q-heterogeneous-sampling}), 
proportion of edges from the original graph or constructed (Section \ref{sec:link-sampling}), 
and selecting edges to refine the overall homophily of the neighborhood (Section \ref{sec:sensitive-attribute-sampling}). 
We denote this entire randomized selection process as a function $\Pi$ (Alg.~\ref{alg:framework}) that outputs the final fair neighborhood $F_i'$ used during training:
\begin{align} \Pi(F_i) = F_i'\end{align}

\subsubsection{Non-observed Homophily Randomization}

\label{sec:q-heterogeneous-sampling}

In the first step, we aim to control the \textit{heterogeneity of non-observed edges} which are added to the training graph. For node $i$, our method constructs edges to $i$ that may be either homogeneous or heterogeneous. Ideally, constructed edges of both types should have been selected such that they increase fairness, but some ratio of homogeneous to heterogeneous may be more optimal. 
Consider the fair neighborhood $F_i$ of $i$ that is created in the rewiring process (Section~\ref{sec:rewiring}). 
Recall that $F_i\setminus{N_i}$ is the set of nodes, which form non-observed edges with $i$ but are added to the neighborhood in  the rewiring process. 
In this first step, we select from these two sets, to form a new set $Q_i$. For some parameter $\alpha \in [0, 1]$, 
we construct the set $Q_i = Q_i^{s_i} \cup Q_i^{\overline{s_i}}$, where each set is defined formally as
\begin{align} 
    \vp &\sim \text{UniformRandom}(0,1) \in \RR^{|F_i\setminus{N_i}|} \\ 
    Q_i^{s_i} &= \big\{ j \in (F_i\setminus{N_i}) \;|\; p_j< \alpha \wedge s_i=s_j \big\} \label{eq:q_i}
    \\
    Q_i^{\overline{s_i}} &= \big\{ j \in (F_i\setminus{N_i}) \;|\; p_j \geq \alpha \wedge s_i\not=s_j \big\}\label{eq:q_i_not}
\end{align}
where probability $p_j$ is sampled uniformly at random for each $j$.
$Q_i^{s_i}$ is the newly constructed edges to nodes with the same sensitive attribute value as $i$, and $Q_i^{\overline{s_i}}$ is the subset of the fair neighborhood with a different sensitive attribute value.
If $\alpha = 1$, then we accept all non-observed homogeneous edges and none of the non-observed heterogeneous edges, and vice versa when $\alpha = 0$.

\subsubsection{Link Randomization}
\label{sec:link-sampling}
Now that we have our set $Q_i$ of constructed edges, which are by definition edges not in the original graph, we would like control the proportion of constructed edges versus original edges in $N_i$.
For parameter $\beta \in [0,1]$, we construct the set $D_i = D_i^n \cup D_i^q$ as follows
\begin{align} 
    \vp &\sim \text{UniformRandom}(0,1) \in \RR^{|N_i\cup Q_i|} \\ 
    D_i^{n} &= \big\{ j \in (N_i\cup Q_i) \;|\; p_j< \beta \wedge j \in N_i \big\} \label{eq:d_n}
    \\
    D_i^{q} &= \big\{ j \in (N_i\cup Q_i) \;|\; p_j \geq \beta \wedge j\in Q_i \big\}\label{eq:d_q} 
\end{align}
\noindent
where probability $p_j$ is sampled uniformly for each. When $\beta = 1$, $D_i = N_i$, i.e. we only select the original positive edges from the graph; when $\beta = 0$, $D_i = Q_i$, i.e. we only select the new edges constructed; and when $\beta = 0.5$, there are exactly the same number of original edges as constructed edges. This enables the user to choose with greater discretion which edges are used in training.

\subsubsection{Sensitive Attribute Randomization}
\label{sec:sensitive-attribute-sampling}
The set $D_i$ now has some combination of our original positive edges and constructed edges. In our final randomization step, we control the level of overall homophily. Let $D_i^{s_i}$ denote nodes with the same sensitive attribute value as $i$, and let $D_i^{\overline{s_i}}$ denote nodes with a different sensitive attribute value. We construct a final set, $F_i'$, where homophily is controlled by parameter $\delta \in [0,1]$, and $F_i' = F_i'^{s_i} \cup F_i'^{\overline{s_i}}$, where $ F_i'^{s_i}$ and  $F_i'^{\overline{s_i}}$ are constructed as followed:
\begin{align} 
    \vp &\sim \text{UniformRandom}(0,1) \in \RR^{|D_i|} \\ 
    F_i'^{s_i} &= \big\{ j \in (D_i) \;|\; p_j< \delta \wedge j \in D_i^{\overline{s_i}} \big\} \label{eq:f_i}
    \\
    F_i'^{\overline{s_i}} &= \big\{ j \in (D_i) \;|\; p_j \geq \delta \wedge j\in  D_i^{s_i} \big\}\label{eq:f_not_i} 
\end{align}
\noindent When $\delta = 1$, we remove all homogeneous edges, keeping only heterogeneous ones; when $\delta = 0$, we remove all heterogeneous edges, keeping only homogeneous ones.

\subsection{Training}\label{sec:training}

Given our fair neighborhood $F_i'$ from Section~\ref{sec:rewiring}-\ref{sec:neigh-selection-randomization}, we then leverage this neighborhood to update the embedding $\vh_i$ for node $i$. In this section, we describe an approach for how these fair neighborhoods are used in training. This approach can be generalized to any existing (or future) state-of-the-art GNN model, since they are reliant on neighborhood updates. 
We do the following for each $k \in [1 \ldots K]$, where $K$ is our search depth:
\begin{align}
     \vh_{F_i^{\prime}}^{k} = \textsc{Aggr}_k\Big(\big\{\vh_j^{k-1},\;\, \forall j \in F_i^{\prime}\big\}\Big) \\
    \vh_i^{k} \leftarrow \sigma\Big(\mW^k \cdot \Psi\Big(\vh_{F_i^{\prime}}^{k}, \vh_i^{k-1}\Big)\Big)
\end{align}
\noindent where $\textsc{Aggr}_k$ is some differentiable aggregator function over the node embeddings in the fair neighborhood $F_i^{\prime}$, $\sigma$ is the non-linear activation function, $\mathbf{W}^k$ is the weight matrix for that iteration, and $\Psi$ is some function combining the representation of the nodes in $i$'s fair neighborhood with the node's current representation, $\vh_i^{k-1}$. In other words, we first aggregate the representations of the fair neighborhood $F_i$ of node $i$ into a vector $\vh_{F_i'}^k$, which is then fed into a fully connected layer and activation function to compute the representation at the next step, $\vh_i^k$. Thus, our framework can be flexibly used in any GNN model, since the fair neighborhoods generated by our framework can be used to create node representations as shown above. Next, we show how any objective function can be leveraged to train on these edges.

\subsubsection{Fair Link Prediction}
The overall objective function for fair link prediction is:
\begin{align}
    \mathcal{L} = \frac{1}{E'}\sum_{(i,j) \in E'} y_{ij} \cdot \log(p(y_{ij}) + (1 - y_{ij})) \cdot \log(1 - p(y_{ij}))
\end{align}
\noindent where $y_{ij}$ is the label $1$ for an existing link and $0$ for no link, and $p(y_{ij})$ is the predicted probability of the edge $ij$ being a positive edge, based on a model trained on our fair embeddings.
Further, 
\begin{align}
    E^{\prime} = F_1' \cup F_2' \cup \cdots \cup F_n' \cup E_{\rm neg}
\end{align}
\noindent
where $E^{\prime}$ consists of observed edges $E$, non-observed constructed edges included to ensure fairness among the neighborhoods
(as well as to improve accuracy of the downstream task, e.g., link prediction), and negative edges denoted by $E_{\rm neg}$, which are sampled uniformly from the set of vertex pairs that do not have an edge between them in the original graph, such that $|E_{\rm neg}| = \frac{|E|}{2}$. 
Note that for the link prediction task, when the rewiring algorithm described in Section \ref{sec:rewiring-task-link-neigh-fairness} 
is used, the constructed, non-observed edges in $ F_1' \cup F_2' \cup \cdots \cup F_n' \setminus E$ are likely to overlap with the test set $E_{\rm test}$; that is, edges added by our rewiring method are selected based on the edges used in training, such that with high probability they are actually observed edges in the graph.

\subsubsection{Fair Link Classification}
We also investigate using the neighborhood fairness framework for fair link classification.
To the best of our knowledge, our work is the first to study fairness in the link classification setting.
More formally, given a graph $G$ along with labels $Y^e_{\rm train}$ for a few of the links (edges), the semi-supervised link classification problem is to predict the remaining link labels.
Hence, let $Y^e$ denote the labels of the links in $G$, then $Y^e_{\rm test} = Y^e \setminus Y^e_{\rm train}$ is the remaining link labels to predict where $|Y^e_{\rm test}| \gg |Y^e_{\rm train}|$.

The proposed neighborhood fairness framework can naturally be leveraged for fair link classification.
Given a link $(i,j) \in E$ with link label $y_{ij} \in Y^e_{\rm train}$, 
we first apply an edge embedding function $\Phi$ to obtain an edge embedding $\vh_{ij}$ as
\begin{align}
    \Phi : \RR^{d} \times \RR^{d} \rightarrow \RR^{d'}
\end{align}
Hence, given $\vh_i \in \RR^{d}$ and $\vh_j \in \RR^{d}$, we derive $\vh_{ij} \in \RR^{d'}$ as
\begin{align}
\vh_{ij} = \Phi(\vh_i,\vh_j)
\end{align}
where $\vh_{ij}$ is a $d'$-dimensional edge embedding and $\Phi$ is an edge embedding function.
Note $\Phi$ can be a concatenation function $\vh_{ij}=[\,\vh_i \; \vh_j\,]$ where $d'=2d$, or a function that combines $\vh_i$ and $\vh_j$ to obtain another $d$-dimensional embedding such as $\vh_{ij}=\vh_i + \vh_j$, $\vh_{ij}= \frac{1}{2}\vh_i + \vh_j$,  or $\vh_{ij}=\vh_i \otimes \vh_j$.
Given $\vh_{ij}$, we then have
\begin{align}
\hat{\vy}^{(ij)} = g(\vh_{ij})
\end{align}
where $\hat{\vy}^{(ij)}$ is the predicted vector of probabilities of the link labels for link $(i,j)$.
Further,
$g : \RR^{d'} \rightarrow \RR^{|Y|}$
where $|Y|$ is the number of unique class labels.
Then a loss such as cross-entropy is used:
\begin{align}\label{eq:link-class-loss}
    \mathcal{L} = -\frac{1}{|Y^e_{\rm train}|}\sum_{(i,j)\in Y^e_{\rm train}}\sum_{k=1}^{|Y|} \vy^{(ij)}_{k} \log \; \hat{\vy}^{(ij)}_{k}
\end{align}
where $\vy^{(ij)}_{k}$ corresponds to the $k$-th element of the one-hot encoded label for link $(i,j)$, that is $\vy^{(ij)} \in \{0,1\}^{|Y|}$ such that $\mathbf{1}^{\top}\vy^{(ij)}=1$, and $\hat{\vy}^{(ij)}_{k}$ is the predicted probability of the $k$-th link label (or more generally weight) from the model.

\section{Experiments}\label{sec:exp}
We design experiments to investigate these research questions: 
\begin{itemize}
    \item \textbf{RQ1:} Is our neighborhood fairness framework and GNN approaches that leverage it able to achieve better fairness while maintaining a similar or even better accuracy
    on link prediction (Section~\ref{sec:exp-fair-link-pred})?

    \item \textbf{RQ2:} For the newly studied problem of \emph{fair link classification}, does our approach lead to better fairness while maintaining a similar or better accuracy (Section~\ref{sec:exp-fair-link-classification})?

    \item \textbf{RQ3:} Does our approach learn fair graph embeddings in general (Section~\ref{sec:exp-fair-embeddings})?
  
\end{itemize}

\noindent In this work, we leverage a wide variety of graph datasets for evaluation.
See Table~\ref{tab:dataset-statistics} for a summary of datasets and their statistics.

\subsection{Experimental Setting}\label{sec:exp-setup}

To evaluate the impact on fairness and accuracy of our approach, we evaluate the output embeddings on downstream tasks of link prediction and node classification. Unless otherwise stated, we use the link prediction rewiring algorithm described in Section \ref{sec:rewiring-task-link-neigh-fairness}. We compute these embeddings on six real-world graphs from various contexts, summarized in Table \ref{tab:dataset-statistics}. We test a GCN with two layers and an embedding dimension of 128, trained with an Adam optimizer for 100 epochs, with a learning rate of 0.005. We perform these experiments 10 times each with different random seeds and with an 80\%/20\% train-test split, and report averages with standard deviations. 
We compare our results against several baselines: GCN, GAT, FairDrop \cite{DBLP:journals/corr/abs-2104-14210}, and FairAdj \cite{FairAdj}. We evaluate each of them based on their accuracy, measured by AUC, and several fairness metrics. 
For link prediction and link classification, we measure the \textit{demographic parity} (DP) and \textit{equalized odds}  (EO) using the implementations from FairDrop~\cite{FairDrop}.

\begin{table}[t]
    \centering
    \caption{
    Link Prediction Results (with input feature setting). Arrows indicate whether larger or smaller values are desired.
    }
    \label{table:link-pred-results-with-input-features}
    \vspace{-3.5mm}
    \begin{tabular}{@{}ll c cc @{}}
    \toprule
    \multicolumn{1}{l}{\textbf{}}   &
    & \multicolumn{1}{c}{\textsc{Accuracy}} 
    & \multicolumn{2}{c}{\textsc{Fairness}} 
    \\
    \cmidrule(lr){3-3}
    \cmidrule(lr){4-5}
    \textbf{Graph} &
    \textbf{Model}         
    & \textbf{AUC}  $\uparrow$  
    & \textbf{DP}   $\downarrow$  
    & \textbf{EO} $\downarrow$ 

    \\
    \midrule
    \multirow{5}{*}{Cora} & 
    GCN  & 0.839 $\pm$ 0.03 &  
    37.01 $\pm$ 3.48 & 
    29.18 $\pm$ 4.96 \\ 
    
    & GAT & 0.859 $\pm$ 0.01 & 
    40.84 $\pm$ 2.70 & 
    32.53 $\pm$ 4.92 
    \\ 
    & FairDrop & 
    0.905 $\pm$ 0.01  & 
    43.24 $\pm$ 1.98 & 
    34.79 $\pm$ 4.49   
    \\
    & FairAdj & 0.840 $\pm$ 0.01 & 
    31.76 $\pm$ 2.43 & 
    10.48 $\pm$ 2.24 
    \\
    \cmidrule(lr){2-5}

    & \textsc{FairNeigh} & 
    0.900 $\pm$ 0.01  & 
    40.94 $\pm$ 1.67 & 
    30.98 $\pm$ 3.35  \\

    \midrule
    \multirow{5}{*}{Citeseer} & 
    GCN  & 
    0.842 $\pm$ 0.02 & 
    36.962 $\pm$ 3.09 & 
    29.01 $\pm$ 4.13 
    \\ 
    & GAT & 
    0.864 $\pm$ 0.01 & 
    41.407 $\pm$ 2.16 & 
    32.78 $\pm$ 4.52
    \\ 
    & FairDrop &  
    0.902 $\pm$ 0.01 &  
    44.17 $\pm$ 2.19 & 
    36.87 $\pm$ 3.95 
    \\ 
    & FairAdj & 
    0.790 $\pm$ 0.02 & 
    13.72 $\pm$ 0.88 & 
    6.61 $\pm$ 1.82 \\
    
\cmidrule(lr){2-5}

& \textsc{FairNeigh}
 &  0.894 $\pm$ 0.01 &  
 38.76 $\pm$ 1.23 & 
 25.94 $\pm$ 3.30
\\ 

    \bottomrule
    \end{tabular}
    \vspace{-2mm}
\end{table}

\begin{table}[h!]
    \centering
    \caption{
    Link Prediction Results for Feature-less Setting.
    }
    \label{table:link-pred-results-feature-less-setting}
    \vspace{-3.5mm}

    \begin{tabular}{@{} ll c cc @{}}
    \toprule
    &
    \multicolumn{1}{l}{\textbf{}}   
    & \multicolumn{1}{c}{\textsc{Accuracy}} 
    & \multicolumn{2}{c}{\textsc{Fairness}} 
    \\
    \cmidrule(lr){3-3}
    \cmidrule(l){4-5}
    \textbf{Graph} &
    \textbf{Model}        
    & \textbf{AUC}  $\uparrow$  
    & \textbf{DP}   $\downarrow$  
    & \textbf{EO} $\downarrow$ 
    \\
    \midrule
    
\multirow{5}{*}{FB-Gender} &
    GCN &  
    0.709 $\pm$ 
    0.00 &  0.54 $\pm$ 
    0.54 & 1.12 $\pm$ 0.52 
    \\ 
    &GAT & 
    0.668 $\pm$ 0.01 &  
    0.70 $\pm$ 0.56 & 
    1.36 $\pm$ 0.91 
    \\
    &FairDrop &  
    0.732 $\pm$ 0.00 &  
    0.34 $\pm$ 0.37 & 
    0.72 $\pm$ 0.42  
    \\
    &FairAdj &   
    0.559 $\pm$ 0.01 &  
    2.18 $\pm$ 0.58 & 
    2.80 $\pm$ 0.73 
    \\
\cmidrule(lr){2-5}
  &\textsc{FairNeigh} &   
  0.754 $\pm$ 0.01 & 
  0.09 $\pm$ 0.11 & 
  0.36 $\pm$ 0.27 
\\

\midrule
\multirow{5}{*}{Retweet} 
& \textsc{GCN} & 
0.716 $\pm$ 0.01 &
47.42 $\pm$ 1.91 &
45.93 $\pm$ 3.57

\\

& \textsc{GAT} & 
0.696 $\pm$ 0.01 &
24.00 $\pm$ 0.72 &
14.56 $\pm$ 3.40 
\\

& FairDrop &
0.579 $\pm$ 0.01 &

19.81 $\pm$ 0.62 &
7.75 $\pm$ 1.69 
\\

& FairAdj & 
0.577 $\pm$ 0.01 & 
5.37 $\pm$ 0.61 & 
8.29 $\pm$ 1.96 
\\

\cmidrule{2-5}

& \textsc{FairNeigh} &
0.730 $\pm$ 0.01 &

29.20 $\pm$ 0.81 &
5.73 $\pm$ 2.85 
\\

\midrule

\multirow{5}{*}{WebKB} 
&\textsc{GCN} &
0.567 $\pm$ 0.04 &

8.22 $\pm$ 4.90 &
19.00 $\pm$ 7.73 
\\

&\textsc{GAT} &

0.543 $\pm$ 0.04 &

8.48 $\pm$ 5.46 &
19.08 $\pm$ 9.28 
\\

&\textsc{FairDrop} & 

0.478 $\pm$ 0.04 &

8.50 $\pm$ 5.77 &
17.47 $\pm$ 9.27 
\\

&FairAdj & 
0.475 $\pm$ 0.02 & 
20.78 $\pm$ 7.11 & 
39.00 $\pm$ 17.00 \\

\cmidrule(lr){2-5}

&\textsc{FairNeigh} & 
0.634 $\pm$ 0.04 &
5.77 $\pm$ 4.29 &
17.16 $\pm$ 8.02 
\\

\midrule

\multirow{5}{*}{Gene} 

&\textsc{GCN} & 

0.646 $\pm$ 0.02 &

17.67 $\pm$ 3.58 &
24.30 $\pm$ 7.03 
\\

&\textsc{GAT} &

0.618 $\pm$ 0.02 &

15.22 $\pm$ 4.77 &
23.37 $\pm$ 6.00
\\

&FairDrop &
0.608 $\pm$ 0.03 &
14.88 $\pm$ 4.85 &
19.91 $\pm$ 6.72 
\\

&FairAdj & 
0.634 $\pm$ 0.03 &
10.56 $\pm$ 3.53 & 
20.75 $\pm$ 3.85 
\\

\cmidrule(lr){2-5}

&\textsc{FairNeigh} & 
0.652 $\pm$ 0.03 &

11.08 $\pm$ 2.96 &
9.42 $\pm$ 4.39
\\

    \bottomrule
    \end{tabular}
\end{table}

\begin{table}[h!]
    \centering
    \caption{
    Additional Fairness Results.
    }
    \label{table:link-pred-results-other-fairness-metrics}
    \vspace{-3.5mm}
    \begin{tabular}{@{}l cc cc @{}}
    \toprule
    \textit{\textbf{Model}}         

    & \textbf{DP$_{g}$}   $\downarrow$ 
    & \textbf{EO$_{g}$} $\downarrow$ 
    & \textbf{DP$_{s}$}   $\downarrow$   
    & \textbf{EO$_{s}$} $\downarrow$ 
    \\
    \midrule
    
    GCN & 
    0.91 $\pm$ 0.44 & 
    1.04 $\pm$ 0.42 & 
    2.78 $\pm$ 0.76 & 
    2.56 $\pm$ 0.86 
    \\ 
    GAT & 
    0.99 $\pm$ 0.57 & 
    0.96 $\pm$ 0.51 & 
    2.85 $\pm$ 0.88 & 
    1.98 $\pm$ 1.03
    \\
    FairDrop & 
    0.42 $\pm$ 0.22 &
    0.58 $\pm$ 0.29 & 
    1.07 $\pm$ 0.46 & 
    1.33 $\pm$ 0.57
    \\
    FairAdj &
    1.73 $\pm$ 0.53 & 
    2.45 $\pm$ 0.78 & 
    2.69 $\pm$ 0.83 & 
    4.15 $\pm$ 1.13 \\
\midrule
  \textsc{FairNeigh} &  
  0.20 $\pm$ 0.10 & 
  0.30 $\pm$ 0.10 & 
  0.42 $\pm$ 0.20 & 
  0.66 $\pm$ 0.21 
\\
    \bottomrule
    \end{tabular}
\end{table}

\begin{table}[b]
    \centering
    \caption{
    Varying the GNN model used in our Fair Neighborhood Framework (\textsc{FairNeigh}-GAT).
    }
    \label{table:varying-GNN-model-in-FairNeigh}
    \vspace{-2.5mm}
    \begin{tabular}{@{} l c cc @{}}
    \toprule
    
    \multicolumn{1}{l}{\textbf{}}   
    & \multicolumn{1}{c}{\textsc{Accuracy}} 
    & \multicolumn{2}{c}{\textsc{Fairness}} 
    \\
    \cmidrule(lr){2-2}
    \cmidrule(l){3-4}
    \textbf{Graph}       
    & \textbf{AUC}  $\uparrow$  
    & \textbf{DP}   $\downarrow$  
    & \textbf{EO} $\downarrow$ 
    \\
    \midrule

\multirow{1}{*}{FB-Gender} & 
0.804 $\pm$ 0.01 &
1.28 $\pm$ 0.21 &
0.22 $\pm$ 0.02 
\\

\multirow{1}{*}{Retweet} &
0.717 $\pm$ 0.01 &
24.45 $\pm$ 1.58 &
6.68 $\pm$ 3.43 
\\

\multirow{1}{*}{WebKB} &
0.637 $\pm$ 0.04 &
5.22 $\pm$ 3.53 &
13.08 $\pm$ 9.21 
\\

Gene & 
0.683 $\pm$ 0.02 &
11.50 $\pm$ 3.02 &
7.59 $\pm$ 4.13 
\\

    \bottomrule
    \end{tabular}
\end{table}

\subsection{Fair Link Prediction Results} \label{sec:exp-fair-link-pred}
To answer RQ1, we evaluate our approach on link prediction (Tables \ref{table:link-pred-results-with-input-features} and \ref{table:link-pred-results-feature-less-setting}).
Compared to the other models, FairNeigh's mean gain in AUC is 5.3\% over GCN, 7.2\% over GAT, 8.9\% over FairDrop, and 15.5\% over FairAdj. Its mean gain in DP is 108.9\% over GCN, 125.0\% over GAT, 57.1\% over FairDrop, and 401.1\% over FairAdj, and its mean gain in EO is 181.0\% over GCN, 103.8\% over GAT, 50.5\% over FairDrop, and 138.2\% over FairAdj.
Overall, FairNeigh achieves an average gain in AUC over all models of 9.2\%, an average gain in DP of 173.0\%, and average gain in EO of 118.4\%. 
We also investigated four other fairness metrics in Table \ref{table:link-pred-results-other-fairness-metrics} where FairNeigh is shown to also perform the best.
Many other results were included in Appendix and others were removed due to space.

While prior work has mainly explored fairness for graphs with features, we also introduce and explore a new setting, which we term the \emph{feature-less setting}.
In this setting, we do not include any underlying features as input, and instead initialize the feature matrix either uniformly at random or based on the graph structure 
(e.g., using SVD or an unsupervised embedding approach~\cite{https://doi.org/10.48550/arxiv.1607.00653,https://doi.org/10.48550/arxiv.1704.08829}).
This new feature-less setting is important to study since many graphs do not naturally come with input features, and when there are input features, they can often be correlated with the sensitive attribute.
As an aside, one can also study graphs with features under this setting by simply ignoring them.
As such, our evaluation leverages both of these important settings.
Results for the feature-less setting are provided in Table \ref{table:link-pred-results-feature-less-setting}.
In particular, our method outperforms all other models in both accuracy and fairness for a variety of graphs in this setting, achieving the highest accuracy in all cases, and 
improving DP by an average of 51.2\% and EO by an average of 26.2\% over the most accurate baselines.

\begin{table}[]
    \centering
    \caption{
    Link Classification Accuracy and Fairness Results.}
    \label{table:link-classification-results}
    \vspace{-3mm}
    \begin{tabular}{l c cc @{}}
    \toprule
    \multicolumn{1}{l}{\textbf{}}   
    & \multicolumn{1}{c}{\textsc{Accuracy}} 
    & \multicolumn{2}{c}{\textsc{Fairness}} 
    \\
    \cmidrule(lr){2-2}
    \cmidrule(lr){3-4}
    \textit{\textbf{Model}}         
    & \textbf{AUC}  $\uparrow$  
    & \textbf{DP}   $\downarrow$  
    & \textbf{EO} $\downarrow$ 
    \\
    \midrule
\textsc{GCN} & 
0.519 $\pm$ 0.11 &
28.23 $\pm$ 16.7 &
62.00 $\pm$ 35.46 
\\

\textsc{GAT} & 
0.566 $\pm$ 0.11 &
38.69 $\pm$ 28.00 &
84.52 $\pm$ 29.37 
\\

\textsc{FairDrop} &
0.517 $\pm$ 0.01 &
1.68 $\pm$ 1.48 &
80.47 $\pm$ 33.81 
\\
\midrule
\textsc{FairNeigh} & 
0.631 $\pm$ 0.13 &
0.43 $\pm$ 1.05 &
26.02 $\pm$ 12.78 
\\

\bottomrule
\end{tabular}
\vspace{-2mm}
\end{table}

\begin{figure}[h]
\centering
\includegraphics[width=0.7\linewidth]{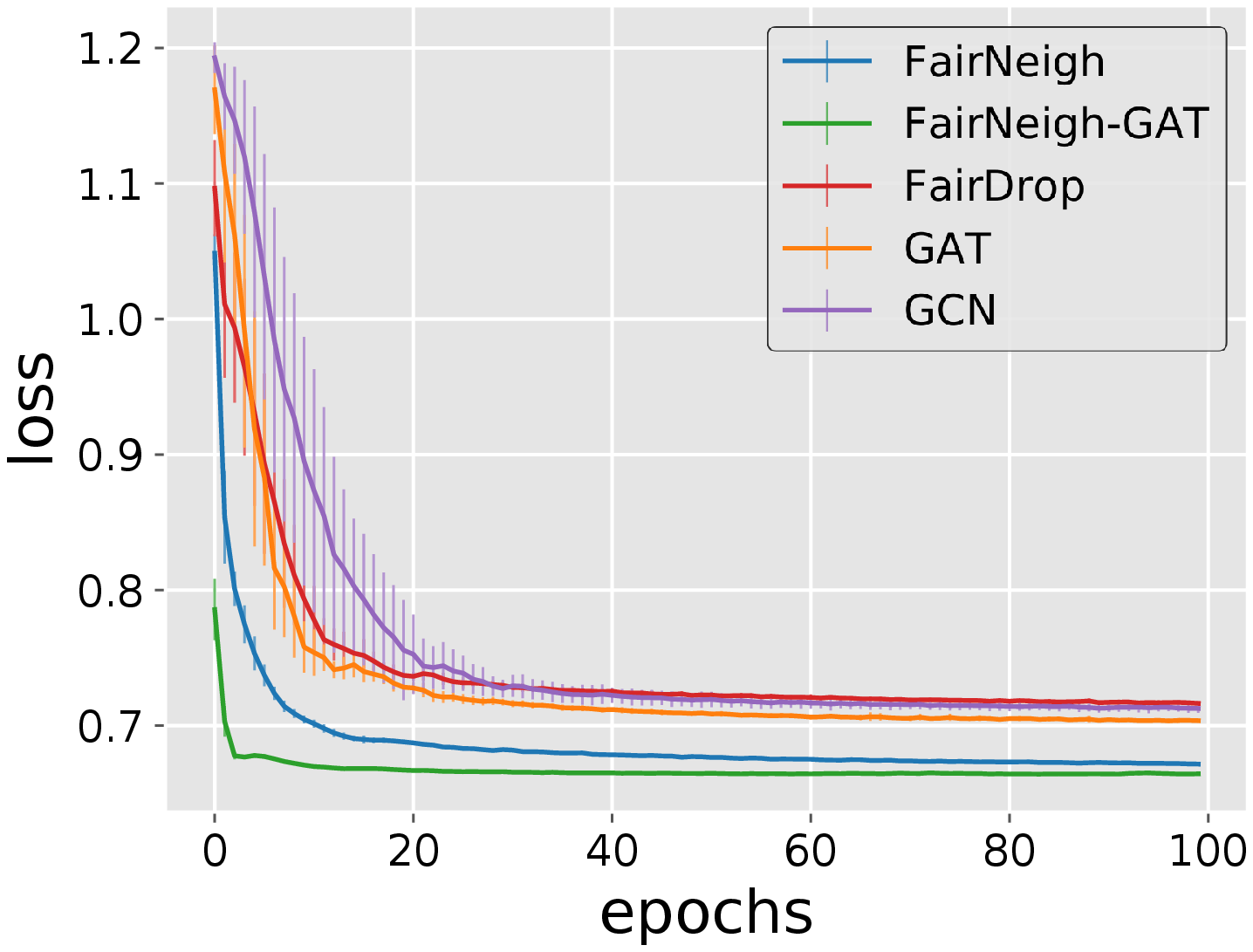}
\vspace{-3mm}
\caption{
Training loss of models from our proposed neighborhood fairness framework compared to other approaches. 
}
\vspace{-2mm}
\label{fig:loss-neigh-fairness-comparison}
\end{figure}

\begin{table*}[t]
\centering
    \caption{
    Representation Bias (RB) results for the setting without features.
    }
    \label{table:rb-results-featureless}
    \vspace{-3.5mm}
    \begin{tabular}{l cc cccccc}
    \toprule
    
    & \multicolumn{2}{c}{\textsc{FB-Gender}} 
    & \multicolumn{2}{c}{\textsc{Retweet}} 
    & \multicolumn{2}{c}{\textsc{WebKB}} 
    & \multicolumn{2}{c}{\textsc{Gene}} 
    \\
    \cmidrule(lr){2-3}
    \cmidrule(lr){4-5}
    \cmidrule(lr){6-7}
    \cmidrule(lr){8-9}
    
    \textbf{Model}         
    & \textbf{AUC}  $\uparrow$   
    & \textbf{RB}   $\downarrow$  
    & \textbf{AUC}  $\uparrow$   
    & \textbf{RB}   $\downarrow$  
    & \textbf{AUC}  $\uparrow$   
    & \textbf{RB}   $\downarrow$  
    & \textbf{AUC}  $\uparrow$   
    & \textbf{RB}   $\downarrow$  
    
    \\ 
    \midrule
GCN  &
0.709 $\pm$ 0.00 & 
0.50 $\pm$ 0.01 & 
0.716 $\pm$ 0.01 & 
0.91 $\pm$ 0.01 & 
0.567 $\pm$ 0.04 & 
0.26 $\pm$ 0.03 & 
0.646 $\pm$ 0.02 & 
0.65 $\pm$ 0.03 
\\ 

GAT  &
0.668 $\pm$ 0.01 & 
0.50 $\pm$ 0.00 & 
0.698 $\pm$ 0.01 & 
0.56 $\pm$ 0.04 & 
0.543 $\pm$ 0.04 & 
0.23 $\pm$ 0.03 & 
0.618 $\pm$ 0.02 & 
0.64 $\pm$ 0.02 
\\ 

FairDrop  &
0.732 $\pm$ 0.00 & 
0.50 $\pm$ 0.00 & 
0.579 $\pm$ 0.01 & 
0.61 $\pm$ 0.01 & 
0.478 $\pm$ 0.04 & 
0.24 $\pm$ 0.03 & 
0.608 $\pm$ 0.03 & 
0.63 $\pm$ 0.01 
\\ 

FairAdj  &
0.559 $\pm$ 0.01 & 
0.50 $\pm$ 0.00 & 
0.577 $\pm$ 0.01 & 
0.51 $\pm$ 0.02 & 
0.475 $\pm$ 0.02 & 
0.20 $\pm$ 0.02 & 
0.634 $\pm$ 0.03 & 
0.56 $\pm$ 0.05 
\\ 
\midrule

FairNeigh  &
0.733 $\pm$ 0.00 & 
0.50 $\pm$ 0.00 & 
0.730 $\pm$ 0.01 & 
0.56 $\pm$ 0.03 & 
0.634 $\pm$ 0.04 & 
0.24 $\pm$ 0.03 & 
0.652 $\pm$ 0.03 & 
0.51 $\pm$ 0.02 
\\ 

\bottomrule
    \end{tabular}
    \vspace{-1mm}
\end{table*}

For the setting with input features, we provide results in Table \ref{table:link-pred-results-with-input-features}.
Notably, FairNeigh achieves the same accuracy as FairDrop, the best-performing baseline, while achieving significantly better fairness.
In particular, FairNeigh has a mean fairness improvement over FairDrop of 8.8\% and 20.3\% in terms of DP and EO, respectively. 
In addition, we investigate using the neighborhood fairness framework to generalize other GNN models such as GAT, which we call \textsc{FairNeigh}-GAT.
Results are provided in Table~\ref{table:varying-GNN-model-in-FairNeigh}.
Strikingly, in nearly all cases, we obtain even better AUC and comparable or sometimes better fairness results when GAT is used as the base GNN model in our fair neighborhood framework.
For instance, \textsc{FairNeigh}-GAT achieves an AUC of 0.804 as shown in Table~\ref{table:varying-GNN-model-in-FairNeigh} compared to 0.754 in Table~\ref{table:link-pred-results-with-input-features} for \textsc{FairNeigh} with GCN.
To understand the data efficiency of our approach compared to the baselines, Figure~\ref{fig:loss-neigh-fairness-comparison} shows the training loss as the number of epochs increases.
It is straightforward to see that \textsc{FairNeigh} and \textsc{FairNeigh}-GAT achieve better data efficiency as both models have consistently lower training loss across all epochs.
For instance, it takes only around 10 epochs to achieve the same loss that the best baseline achieves using 100 epochs.
This significant improvement in data efficiency is due to the careful neighborhood fairness rewiring that chooses edges that are more likely to improve link prediction accuracy. 
As an aside, we observe that GNN models trained using our neighborhood fairness framework have lower standard error and variance compared to the training loss from the other approaches.

\subsection{Fair Link Classification}\label{sec:exp-fair-link-classification}
To answer RQ2, we evaluate our method on link classification.
In this task, we are given a small fraction of link labels for training and need to predict the remaining held-out labels for the links.
In addition, we have a sensitive attribute on the nodes that is never seen by the algorithm.
For the fair link classification task, we use a hospital contact network dataset.
This dataset has 10 link labels representing various interaction types between the various personnel at the hospital.
The sensitive node attribute is the actual roles of the individuals in the hospital.
These include patients, nurses, administrators, and medical doctors.
Edges exist in the graph if two individuals interacted recently or came into close contact 
for an extended period of time.
There are a total of 1139 edges between 75 individuals.
This dataset has a node homophily score of 0.241, and thus is on the heterophilous side.
We are the first to investigate the problem of graph fairness for the link classification task.

Since link classification has not been studied in the context of fairness, there are not any existing methods for direct comparison.
For instance, FairDrop and other approaches are developed specifically for link prediction and as such optimize a loss specific to that setting.
Therefore, we adapted several standard approaches such as GCN and GAT for comparison against our proposed neighborhood fairness framework.
Results are provided in Table~\ref{table:link-classification-results}.
We also provide results for FairDrop using the same link classification loss.
For link classification, we use fairness metrics based on DP and EO.
Strikingly, our approach achieves the best accuracy and fairness across all other methods as shown in Table~\ref{table:link-classification-results}.

\begin{table}[h]
\centering
\vspace{2mm}
    \caption{
    RB results for setting with input features.
    }
    \label{table:rb-results-features}
    \vspace{-3.5mm}
    \begin{tabular}{@{}l cc cc}
    \toprule
    & \multicolumn{2}{c}{\textsc{Cora}} 
    & \multicolumn{2}{c}{\textsc{Citeseer}} 
    \\
    \cmidrule(lr){2-3}
    \cmidrule(lr){4-5}
    
    \textbf{Model}       
    & \textbf{AUC}  $\uparrow$   
    & \textbf{RB}   $\downarrow$      
    & \textbf{AUC}  $\uparrow$   
    & \textbf{RB}   $\downarrow$  
    \\
    \midrule
    GCN & 
    0.839 $\pm$ 0.03  &
    0.79 $\pm$ 0.02 &
    0.842 $\pm$ 0.02  &
    0.78 $\pm$ 0.02
    \\
    GAT & 
    0.859 $\pm$ 0.01 &
    0.78  $\pm$ 0.02 &
    0.864 $\pm$ 0.01 &
    0.78  $\pm$ 0.02
    \\
    FairDrop  & 
    0.905 $\pm$ 0.01 &
    0.80 $\pm$ 0.01 &
     0.902 $\pm$ 0.01 &
    0.80 $\pm$ 0.01
    \\
    FairAdj  & 
    0.840 $\pm$ 0.01 &
    0.69 $\pm$ 0.01 &
     0.790 $\pm$ 0.02 &
    0.31 $\pm$ 0.03
    \\
    \midrule
    FairNeigh  & 
     0.900 $\pm$ 0.01 &
    0.77 $\pm$ 0.02 &
    0.894 $\pm$ 0.01 &
    0.77 $\pm$ 0.02
    \\

\bottomrule
    \end{tabular}
    \vspace{-2mm}
\end{table}

\subsection{Fair Graph Embeddings} \label{sec:exp-fair-embeddings}
In this section, we answer RQ3, which asks whether the proposed neighborhood fairness framework can learn fair graph embeddings such that the node embeddings given as output cannot be leveraged by an adversary to recover the sensitive attribute values of the nodes in the graph.
We use the notion of representation bias to evaluate the fairness of our learned embeddings.
More formally, for classifier $c$, let $P_c(s,\vz_i)$ denote the estimated probability that node $i$ with embedding $\vz_i$ has sensitive attribute value $s \in S$.  
Then representation bias (RB) score~\cite{pmlr-v119-buyl20a} is:
\begin{align}\label{eq:rep-bias}
    \text{RB} = \sum_{s \in S} \frac{1}{V_s} \texttt{AUC}\big(\{P_c(A(j) | \vz_j) | j \in V_s \}\big)
\end{align}
where $V_s = \{j \in V | A(j) = s \}$ and $A(j)$ is the sensitive attribute value for node $j$.
Eq.~\ref{eq:rep-bias} uses weighted one-vs-rest AUC score to measure prediction performance.
Intuitively, if a model learns fair embeddings, then the classifier trained using the node embeddings should perform poorly (close to random if truly independent).
For the setting without features, we report accuracy and representation bias results in Table \ref{table:rb-results-featureless}. 
For all graphs, FairNeigh achieves higher accuracy than all other models, while improving fairness by an average of 14\% over the best performing baseline. 
Compared to the next highest model, which is the GCN for Retweet, WebKB, and Gene, FairNeigh causes an average reduction in representation bias of 24.27\% percent. 
Due to space constraints, we omitted some results, but all were consistent with the findings discussed.
For the setting with input features, the results for representation bias in graphs with features can be found in Table \ref{table:rb-results-features}; alternative columns show accuracy and representation bias with standard deviations side by side for each graph. FairNeigh matches the AUC of the most accurate baseline, FairDrop, while also reducing representation bias by an average of 3.5\%. The only baseline with lower representation bias,  FairAdj, reduces accuracy by almost 10\%. Our approach has the greatest fairness improvement while preserving accuracy.

\section{Conclusion} \label{sec:conc}
This work proposed \textsc{FairNeigh}: a flexible, task-dependent framework for producing fair graph embeddings. Despite the fundamental importance of neighborhoods on the fairness of trained GNN models, \textsc{FairNeigh} is, to the best of our knowledge, the first approach to properly define and address the mitigation of localized neighborhood-level fairness. Our framework first performs intelligent neighborhood rewiring for each graph node to make its neighborhood fair, then performs randomized selection based on parameters set by the user to control levels of homophily and heterophily, as well as the proportion of constructed edges used in training. These fair neighborhoods are then passed through GNN training to produce fair graph embeddings. The experiments demonstrated the effectiveness of our method for improving fairness, and in many cases, accuracy, of a variety of GNN-based tasks, across a variety of application graphs.
To the best of our knowledge, we provided the first investigation into the graph representation learning task of fair link classification. 
Finally, future work should continue to explore the notion of neighborhood fairness and develop more sophisticated learning techniques to ensure fairness across a variety of graph representation learning tasks and datasets.

\balance
\bibliographystyle{ACM-Reference-Format}
\bibliography{paper}

\newpage
\balance
\appendix 
\section*{Appendix}

\section{Fairness accuracy tradeoff}
\begin{figure}[h]
\centering
\includegraphics[width=0.7\linewidth]{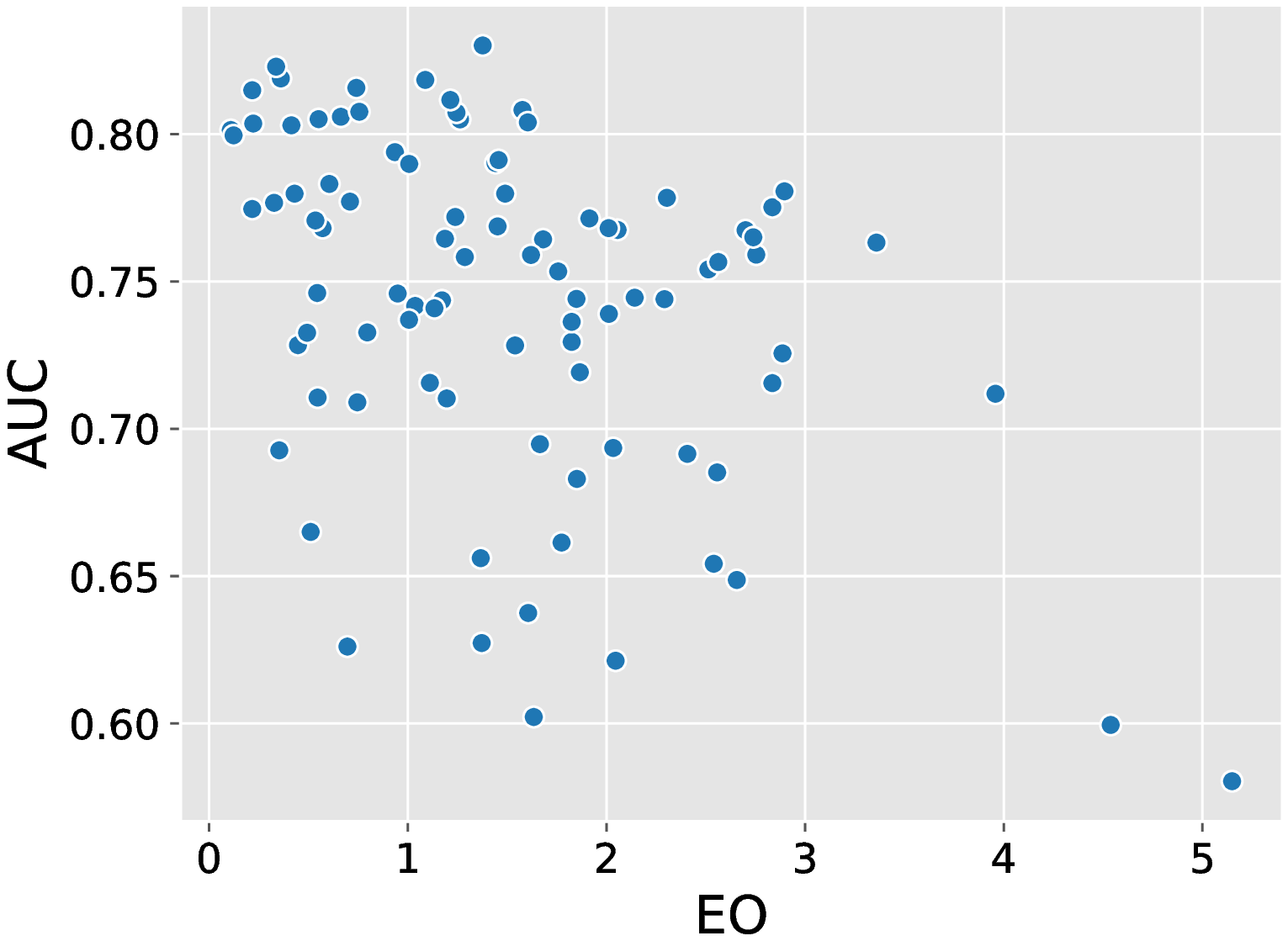}

\vspace{-2mm}
\caption{%
Fairness without Accuracy Loss.
FairNeigh is able to achieve high accuracy and fairness, without having to trade-off one for the other.
}
\label{fig:accuracy-vs-fairness-tradeoff}
\end{figure}

\begin{figure}[b]
\centering
\vspace{-2mm}
\includegraphics[width=0.49\linewidth]{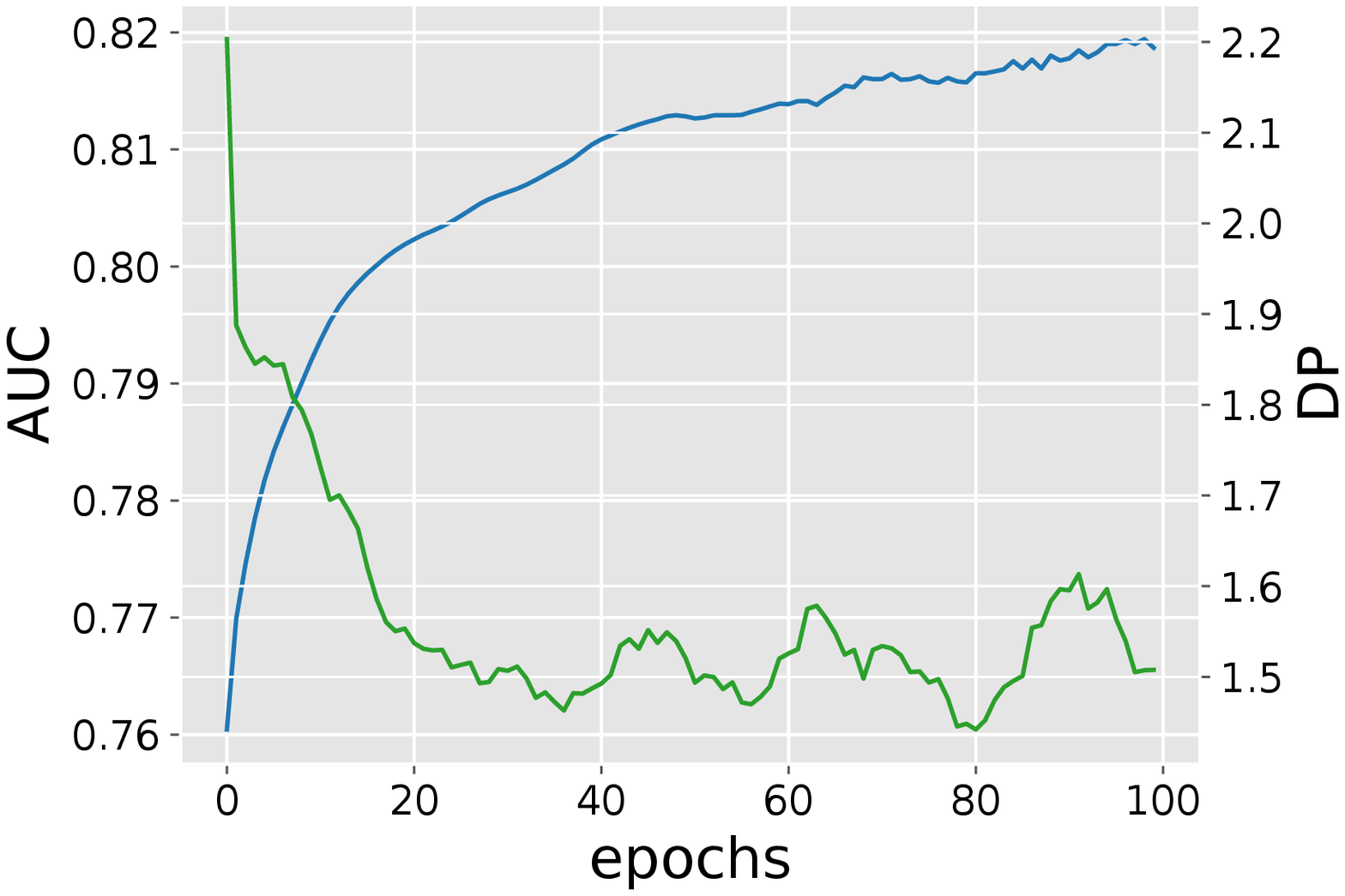} \includegraphics[width=0.49\linewidth]{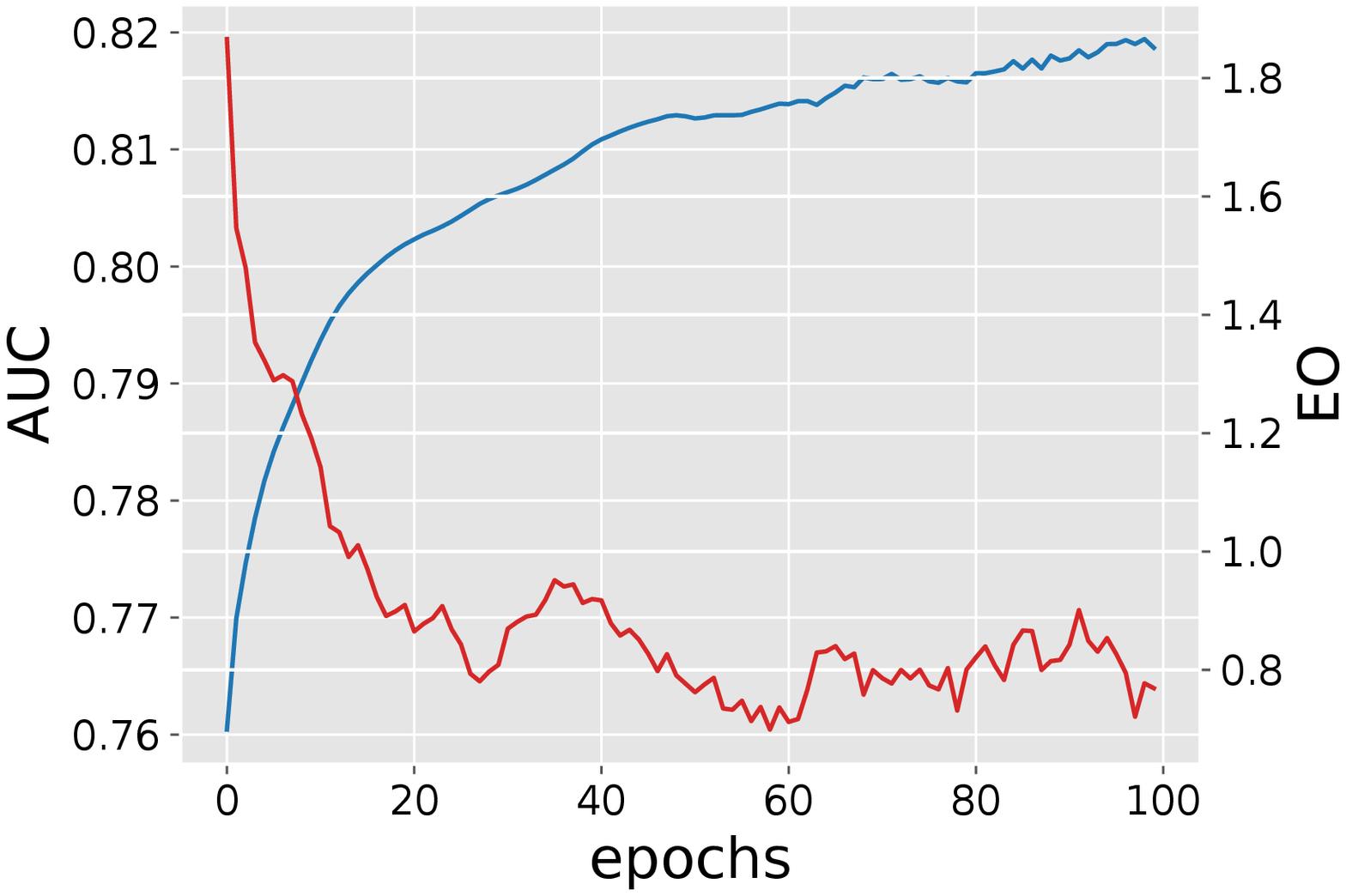}
\vspace{-2mm}
\caption{%
FairNeigh achieves fairness without accuracy loss.
See text for discussion.
}
\label{fig:accuracy-vs-fairness-tradeoff-curves}
\end{figure}

We now analyze the fairness and accuracy trade-off of FairNeigh to understand whether FairNeigh can achieve fairness without having to trade-off accuracy.
We provide results in Figure~\ref{fig:accuracy-vs-fairness-tradeoff} showing the accuracy (AUC) and fairness (EO) for various FairNeigh models trained using different hyperparameters.
For brevity, we use EO as the fairness metric, though similar results were found using others.
Notably, FairNeigh achieves high accuracy and fairness,  without having to significantly trade-off one for the other.
Furthermore, we also analyze fairness and accuracy as the number of epochs used during training increases.
Results are provided in Figure~\ref{fig:accuracy-vs-fairness-tradeoff-curves}. 
It is straightforward to see that accuracy (AUC; higher is better) and fairness (lower is better) are worst when using only a few epochs for training.
However, as the number of epochs increases, the model becomes more accurate while becoming increasingly more fair as well before stabilizing.
Notably, our model achieves good fairness (DP and EO), which stabilizes around 20 epochs, while achieving good accuracy.
Moreover, the accuracy of our fair neighborhood GNN continues to improve as the number of epochs increases, while maintaining fairness (with minor fluctuations).

\begin{table*}[h!]
    \centering
    \caption{
    Link Prediction Results for Gene.
    Best result for each setting is in bold.
    }
    \label{table:link-pred-results-gene}
    \vspace{-2mm}
    \begin{tabular}{l c cc cc cc }
    \toprule
    \multicolumn{1}{l}{\textbf{}}   
    & \multicolumn{1}{c}{\textsc{Accuracy}} 
    & \multicolumn{6}{c}{\textsc{Fairness}} 
    \\
    \cmidrule(lr){2-2}
    \cmidrule(lr){3-8}
    \textit{\textbf{Model}}         
    & \textbf{AUC}  $\uparrow$  
    & \textbf{DP}   $\downarrow$  
    & \textbf{EO} $\downarrow$ 
    & \textbf{DP$_{g}$}   $\downarrow$ 
    & \textbf{EO$_{g}$} $\downarrow$ 
    & \textbf{DP$_{s}$}   $\downarrow$   
    & \textbf{EO$_{s}$} $\downarrow$ 
    \\
    \midrule

    \textsc{GCN} & 
0.646 $\pm$ 0.02 &
17.67 $\pm$ 3.58 &
24.30 $\pm$ 7.03 &
4.15 $\pm$ 1.87 &
6.04 $\pm$ 3.39 &
21.85 $\pm$ 3.19 &
28.15 $\pm$ 6.22 

\\

\textsc{GAT} &
0.618 $\pm$ 0.02 &
15.22 $\pm$ 4.77 &
23.37 $\pm$ 6.00 &
5.52 $\pm$ 1.76 &
7.74 $\pm$ 2.88 &
20.18 $\pm$ 5.26 &
28.12 $\pm$ 4.76 

\\

FairDrop &
0.608 $\pm$ 0.03 &
14.88 $\pm$ 4.85 &
19.91 $\pm$ 6.72 &
8.40 $\pm$ 3.96 &
12.19 $\pm$ 5.17 &
22.96 $\pm$ 6.45 &
28.22 $\pm$ 6.53
\\

FairAdj & 
0.634 $\pm$ 0.03 &
10.56 $\pm$ 3.53 & 20.75 $\pm$ 3.85 & 3.53 $\pm$ 3.30 & 9.61 $\pm$ 2.32 & 14.77 $\pm$ 6.89 & 25.91 $\pm$ 4.09 

\\
\midrule
 
\textsc{FairNeigh} & 
0.652 $\pm$ 0.03 &
11.08 $\pm$ 2.96 &
9.42 $\pm$ 4.39 &
5.79 $\pm$ 3.04 &
9.44 $\pm$ 4.95 &
16.60 $\pm$ 3.08 &
16.32 $\pm$ 5.55 
\\

\textsc{FairNeigh-GAT} &
0.683 $\pm$ 0.02 &
11.50 $\pm$ 3.02 &
7.59 $\pm$ 4.13 &
4.49 $\pm$ 2.75 &
7.57 $\pm$ 4.65 &
15.69 $\pm$ 4.11 &
13.00 $\pm$ 5.48 
\\

\bottomrule
\end{tabular}
\end{table*}

\begin{table}
    \centering
    \caption{
    Representation Bias Results.
    Best result for each setting is in bold.
    }
    \label{table:rep-bias-results-gene}
    \vspace{-2mm}
    \begin{tabular}{l ccc }
    \toprule
    \multicolumn{1}{l}{\textbf{}}   
    & \multicolumn{3}{c}{\textsc{Representation Bias}} 
    \\
    \cmidrule(){2-4}
    \textit{\textbf{Model}}         
    & \textbf{LR} $\downarrow$ 
    & \textbf{MLP} $\downarrow$ 
    & \textbf{RF} $\downarrow$ 
    \\
    \midrule
    
\textsc{GCN} & 
0.606 $\pm$ 0.03 &
0.664 $\pm$ 0.03 &
0.649 $\pm$ 0.03 
\\

\textsc{GAT} &
0.610 $\pm$ 0.03 &
0.662 $\pm$ 0.03 &
0.636 $\pm$ 0.02 
\\

FairDrop &
0.560 $\pm$ 0.03 &
0.635 $\pm$ 0.02 &
0.632 $\pm$ 0.01 
\\

FairAdj & 
0.537 $\pm$ 0.02& 0.622 $\pm$ 0.02& 0.561 $\pm$ 0.05
\\

\midrule

\textsc{FairNeigh} & 
\textbf{0.504} $\pm$ 0.02 &
\textbf{0.532} $\pm$ 0.03 &
\textbf{0.513} $\pm$ 0.02 
\\

\textsc{FairNeigh-GAT} &
0.521 $\pm$ 0.03 &
0.579 $\pm$ 0.02 &
0.557 $\pm$ 0.03 
\\

\bottomrule
\end{tabular}
\end{table}

\section{Neighborhood Fairness Analysis} \label{sec:neigh-fairness-metric}

\begin{figure}[b]
\centering
\hfill
\hspace{-5mm}
\subfigure[Cora]{
\includegraphics[width=0.50\linewidth]{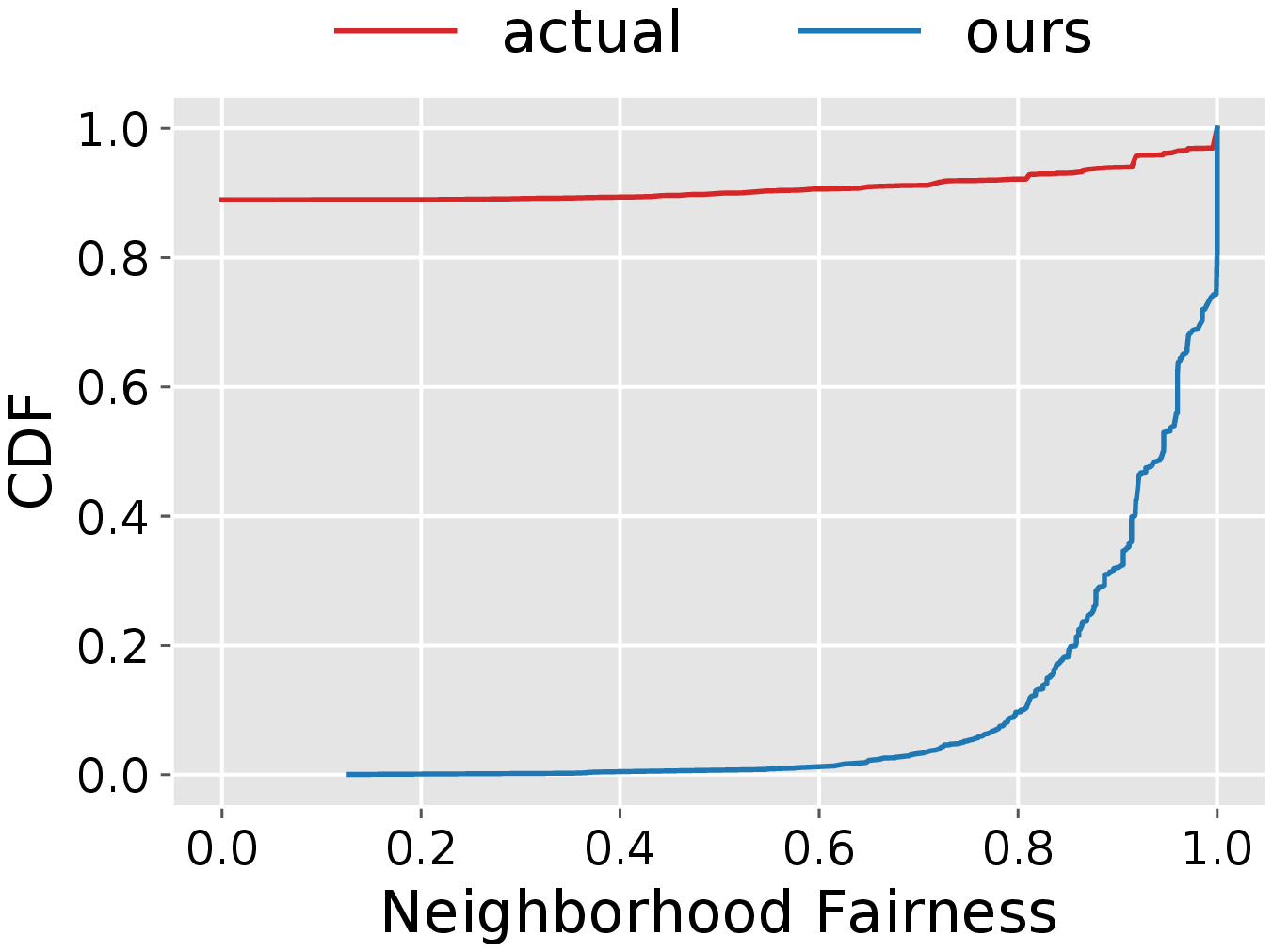}
}
\hfill
\hspace{-5mm}
\subfigure[Citeseer]{
\includegraphics[width=0.50\linewidth]{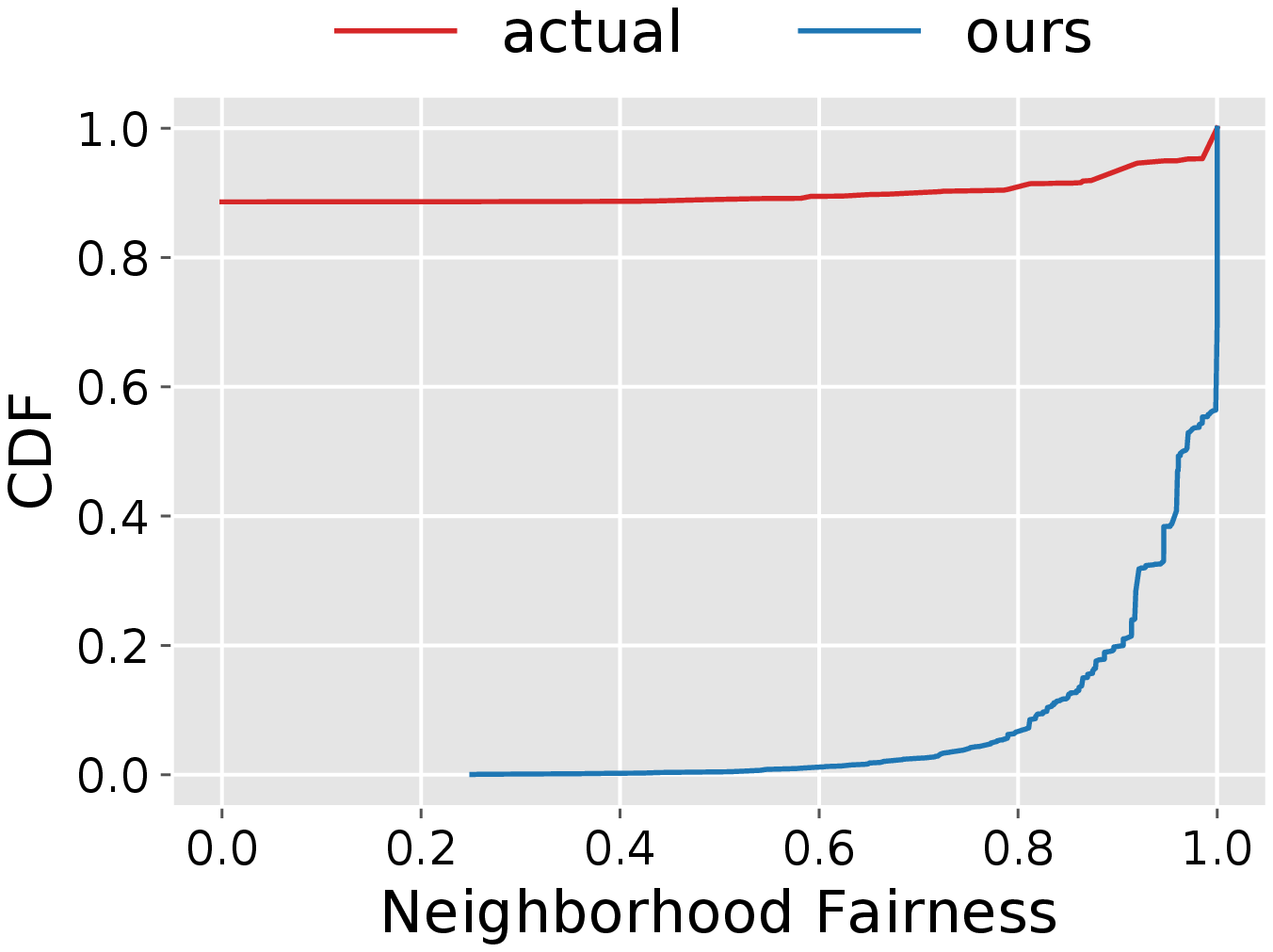}
}

\vspace{-3mm}
\caption{
Neighborhood Fairness Analysis.
Most neighborhoods in real-world graphs are completely \emph{unfair} (\textcolor{red}{actual}).
However, our approach derives fair neighborhoods (\textcolor{blue}{ours}) and leverages these for learning fair GNN models with fair embeddings. 
See text for discussion.
}
\label{fig:neigh-fairness-actual-vs-ours}
\end{figure}

We formally introduce a neighborhood fairness metric that can be leveraged prior to training a graph neural network model to determine the overall localized fairness.
This metric reveals the overall local fairness when a GNN-based approach is used, since these methods all leverage neighborhoods for learning the embeddings of the nodes in the graph.
Therefore, this metric can reveal the overall fairness apriori to training a large-scale GNN model, and based on this, can leverage our approach or future state-of-the-art to mitigate the identified fairness issues that are revealed by the neighborhood fairness metric.
More formally, the entropy-based neighborhood fairness metric is defined as follows:
\begin{Definition}[Local Node Neighborhood Fairness]\label{def:entropy-based-neigh-fairness-metric}
Let $\vc_i$ be the vector of the frequency of the sensitive attribute values of the neighbors $N_i$ of node $i$ such that $c_{ik}=|N_i^{k}|$ where $N_i^{k}$ is the subset of $N_i$ with sensitive attribute value $k$.
Then the neighborhood fairness metric quantifying the localized fairness of a neighborhood of a node $i$ is:
\begin{align}\label{eq:entropy-based-node-neigh-fairness-metric}
    \mathbb{F}(\vp_i) = -\sum_{k} \; p_{ik} \log p_{ik}
\end{align}\noindent
where $\vp_i = \frac{\vc_i}{\sum_k c_{ik}}$ is the probability distribution vector $\sum_k p_{ik}=1$ of node $i$.
Intuitively, when $\mathbb{F}(\vp_i)=1$, then the neighborhood of $i$ is said to be completely fair, as no information is revealed from the neighborhood of $i$ about the sensitive attribute value of $i$.
Hence, $\mathbb{F}(\vp_i)=0$ indicates a neighborhood with minimum fairness (max unfairness) whereas $\mathbb{F}(\vp_i)=1$ indicates a neighborhood with maximum fairness.
\end{Definition}
\noindent Using Definition~\ref{def:entropy-based-neigh-fairness-metric}, we define the overall neighborhood fairness metric of a graph $G$  as follows:
\begin{Definition}[Neighborhood Fairness]\label{def:neigh-fairness-metric-G}
The neighborhood fairness $\mathbb{F}(G)$ of a graph $G$ is 
\begin{align}\label{eq:neigh-fairness-metric-G}
    \mathbb{F}(G) = \frac{1}{|V|} \sum_{i \in V} \mathbb{F}(\vp_i)
\end{align}\noindent
where $\mathbb{F}(G)$ is an intuitive metric characterizing the inherit fairness of $G$ over all the local neighborhoods. 
Thus, capturing the local fairness of the graph $G$ with respect to the sensitive attribute $\vs = \big[\,s_1\, s_2\, \cdots\, s_i\, \cdots\, s_n\,\big]$.
\end{Definition}

We use this neighborhood fairness metric to analyze the fairness of neighborhoods in a variety of real-world graphs.
Strikingly, we observe in Figure~\ref{fig:neigh-fairness-actual-vs-ours} that most neighborhoods in real-world graphs are \emph{unfair} (\textcolor{red}{actual}) since around 90\% of node neighborhoods have the maximum unfairness score $\mathbb{F}(\vp_i)=0$ and almost no neighborhoods are completely fair with $\mathbb{F}(\vp_i)=1$.
In contrast, our approach derives fair neighborhoods and leverages these for learning fair GNN models with fair embeddings for downstream tasks as shown by the blue curve (\textcolor{blue}{ours}).
These results demonstrate that FairNeigh significantly improves the fairness of neighborhoods.

\begin{figure}[h!]
\centering

\hfill
\hspace{-6mm}
\subfigure[AUC]{
\includegraphics[width=0.54\linewidth]{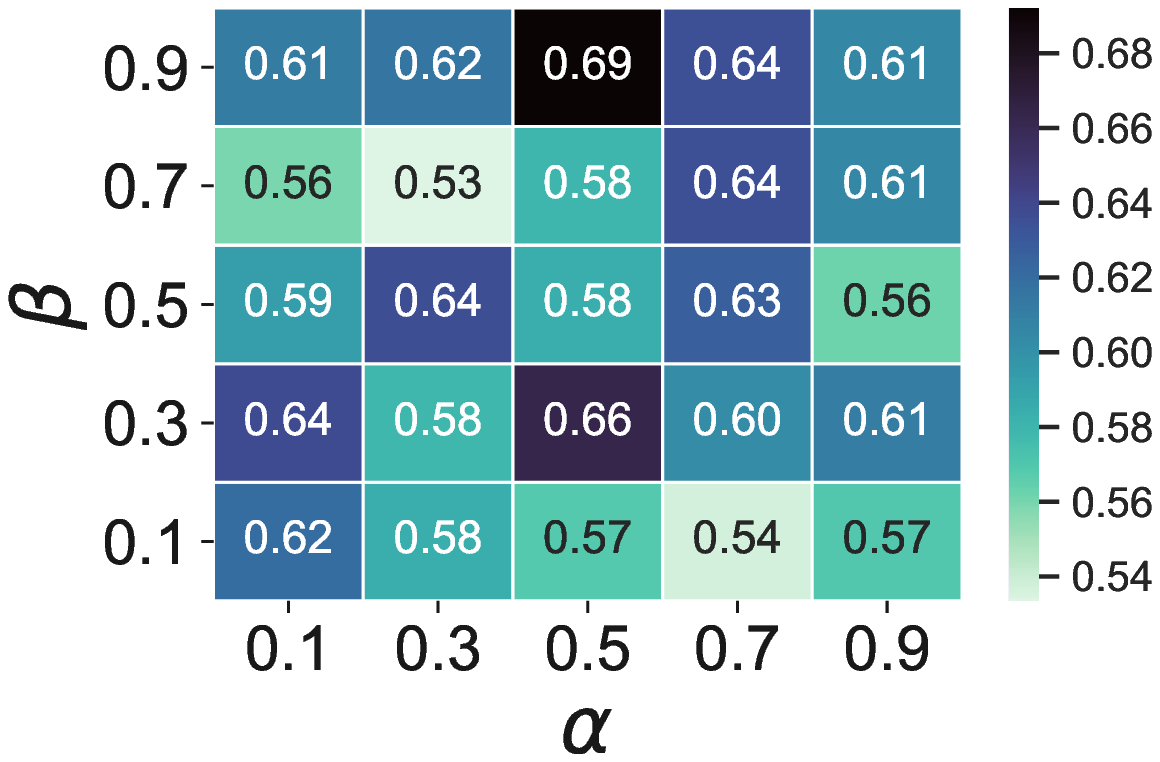}
\label{fig:vary-alpha-vs-beta-AUC}
}
\hspace{-7.0mm}
\subfigure[Fairness (DP)]{
\includegraphics[width=0.54\linewidth]{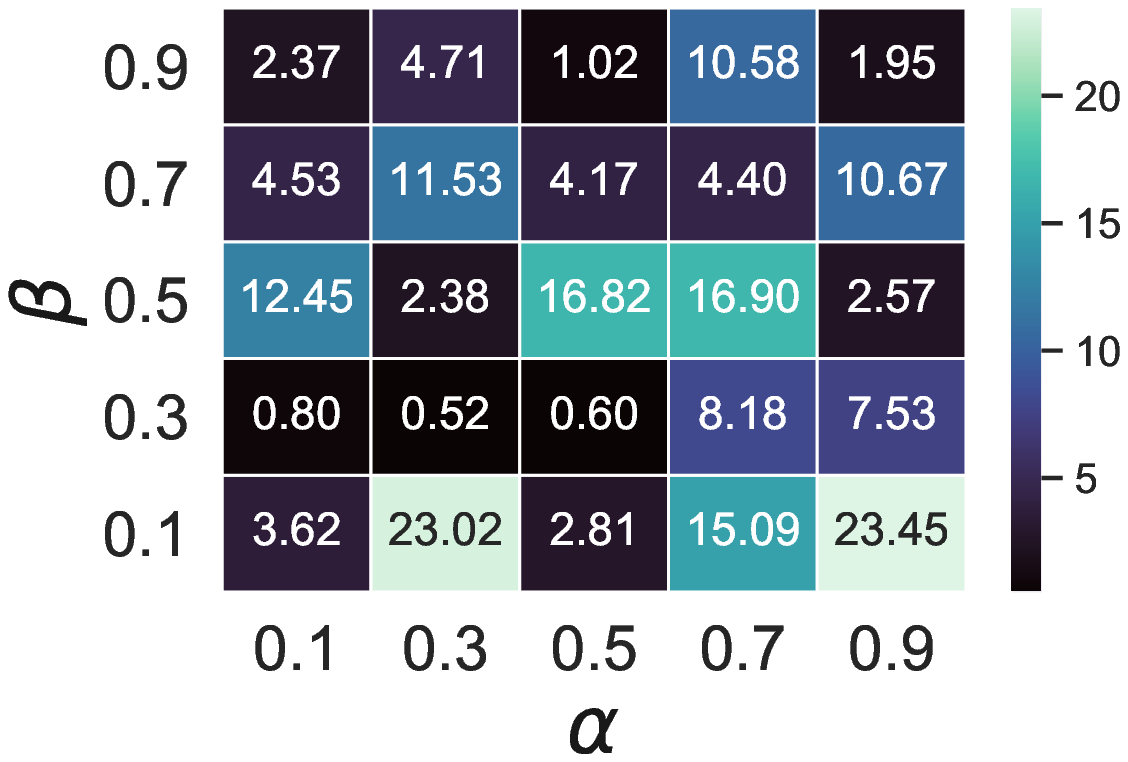}
\label{fig:vary-alpha-vs-beta-DP}
}
\vspace{-2mm}
\caption{%
Accuracy and fairness results as we vary $\alpha, \beta \in \{0.1,0.3,0.5,0.7,0.9\}$.
See text for discussion.
}
\label{fig:vary-alpha-vs-beta}
\end{figure}

\section{Hyperparameter tuning} \label{sec:exp-ablation-study}
Now we investigate the accuracy and fairness as we vary the neighborhood fairness selection randomization hyperparameters used in our fair neighborhood GNN models.
In particular, we vary $\alpha, \beta \in \{0.1,0.3,0.5,0.7,0.9\}$ and measure accuracy and fairness using a fixed $\delta=0.25$. 
Results are provided in Figure~\ref{fig:vary-alpha-vs-beta}.
We observe that FairNeigh achieves the best accuracy when $\alpha=0.5$ and $\beta=0.9$, similarly, the best fairness is achieved when $\alpha=0.3$ and $\beta=0.3$.
However, depending on the importance of accuracy or fairness, one can easily decide how to set such hyperparameters.
Notably, there is an insignificant level of trade-off between accuracy and fairness when $\alpha=0.5$ and $\beta=0.9$, since FairNeigh is able to achieve the best accuracy of 0.69, while achieving a near-optimal level of fairness of 1.02.
Conversely, if $\beta=0.3$, then FairNeigh achieves an AUC of 0.66 (2nd best) while achieving a better fairness of 0.60 as shown in Figure~\ref{fig:vary-alpha-vs-beta}.
This highlights the flexibility of the proposed fair neighborhood framework, and the importance of our fair neighborhood selection randomization.
Due to space constraints some results were removed, however, we observed even better results as $\delta$ varies along with $\alpha$ and $\beta$ as well.

\begin{table}[thb]
    \centering
    \caption{Graph statistics and properties.
    }
    \label{tab:dataset-statistics}
    \vspace{-3mm}
    \begin{tabular}{l@{}cccc}
    \toprule
    & \textbf{\#Sensitive} & & & \\
    \textbf{Dataset} &  
    \textbf{Attr. Values} &
    \textbf{\#Nodes} &  
    \textbf{\#Edges}  & 
    \textbf{Homophily} \\
    \midrule
    
    Retweet-pol. & 2 & 18470 & 61157 &  0.979 \\
    
    Gene & 2 & 1103 & 1672 & 0.833 \\
    
    Cora        & 7     & 2,708     & 5,429     &  0.814 \\
    Citeseer    & 6     & 3,327     & 4,732     &  0.738 \\
  
    FB-Gender & 2 & 7315 & 89733 &  0.616 \\
    WebKB & 5 & 265 & 530 &  0.204 \\
    
    \bottomrule
    \end{tabular}
\end{table}

\section{Additional Fairness Results}
\label{sec:full-tables}
We show several more fairness results in Tables \ref{table:link-pred-results-gene} and \ref{table:rep-bias-results-gene}. In Table \ref{table:link-pred-results-gene}, we present additional link prediction results for gene for four more fairness metrics: group and subgroup demographic parity and equalized odds ($DP_g$, $EO_g$, $DP_s$, $EO_s$), and in Table \ref{table:rep-bias-results-gene}, we include representation bias in gene measured by two other models. In all cases, \textsc{FairNeigh} is able to achieve higher AUC, and in most cases, achieves better (lower) unfairness scores, with lower subgroup unfairness than almost all models. We find similar results for all graphs mentioned in the paper, but omit them to preserve space.

\section{Datasets}
In Table \ref{tab:dataset-statistics}, we provide statistics and properties of the graph datasets used in the experiments.
As shown, we evaluated FairNeigh on a variety of graphs, which ranged in size, homophily, sensitive attributes, and contexts.

\end{document}